\documentclass[dvipsnames]{article}

\PassOptionsToPackage{numbers, compress}{natbib}
\usepackage[preprint]{neurips_2024}

\usepackage[utf8]{inputenc} % allow utf-8 input
\usepackage[T1]{fontenc}    % use 8-bit T1 fonts
\usepackage{hyperref}       % hyperlinks
\usepackage{url}            % simple URL typesetting
\usepackage{booktabs}       % professional-quality tables
\usepackage{longtable}
\usepackage{threeparttable}
\usepackage{tabularx}
\usepackage{float}
\usepackage{amsfonts}       % blackboard math symbols
\usepackage{nicefrac}       % compact symbols for 1/2, etc.
\usepackage{microtype}      % microtypography
\usepackage{xcolor}         % colors
\usepackage{array}
\usepackage{graphicx}
\usepackage{amsmath}
\usepackage{cleveref}
\usepackage{multirow}
\usepackage{subcaption}
\usepackage{algorithm}    % provides the algorithm float
\usepackage{algpseudocode}  % alternative to algorithmic
\usepackage[colorinlistoftodos]{todonotes}
\usepackage{standalone}
\usepackage{placeins}

\usepackage{tikz,pgfplots}
\usetikzlibrary{calc,external,backgrounds,3d,pgfplots.colormaps,positioning}

\newcommand{\colliderml}{\textsc{ColliderML}}
\newcommand{\geant}{\texttt{\textsc{Geant4}}}
\newcommand{\pythia}{\texttt{\textsc{Pythia8}}}
\newcommand{\ddforhep}{\texttt{\textsc{DD4hep}}}
\newcommand{\keyforhep}{\texttt{\textsc{Key4hep}}}
\newcommand{\edmforhep}{\texttt{\textsc{EDM4hep}}}
\newcommand{\acts}{\texttt{\textsc{ACTS}}}
\newcommand{\ttbar}{$t\bar{t}$ }

\title{ColliderML: The First Release of an OpenDataDetector High-Luminosity Physics Benchmark Dataset}

\author{%
  Doğa Elitez \\ 
  CERN \&  Johannes \\ Gutenberg-Universität Mainz \\
  \And
  Paul Gessinger \\
  CERN \\
  \And
  Daniel Murnane\thanks{Corresponding author: \texttt{daniel.murnane@nbi.ku.dk}} \\
  Niels Bohr Institute \\ \& Lawrence Berkeley National Lab \\
  \And 
  Marcus Selchou Raaholt \\
  University of Copenhagen
  \And
  Andreas Salzburger \\
  CERN \\
  \And 
  Stine Kofoed Skov \\
  University of Copenhagen \\
  \And
  Andreas Stefl \\
  CERN \\
  \And
  Anna Zaborowska \\
  CERN
}

\begin{document}

\maketitle
\setcounter{footnote}{0}

\begin{abstract}
  \noindent We introduce \textbf{\colliderml}\footnote{Documentation: \url{https://opendatadetector.github.io/ColliderML/}, \\ Dataset: \url{https://huggingface.co/datasets/CERN/ColliderML-Release-1}} - a large, open, experiment-agnostic dataset of fully simulated and digitised proton--proton collisions in High-Luminosity Large Hadron Collider conditions (\(\sqrt{s}=14\,\mathrm{TeV}\), mean pile-up \(\mu\ = 200\)). \colliderml{} provides one million events across ten Standard Model and Beyond Standard Model processes, plus extensive single-particle samples, all produced with modern next-to-leading order matrix element calculation and showering, realistic per-event pile-up overlay, a validated OpenDataDetector geometry, and standard reconstructions. The release fills a major gap for machine learning (ML) research on detector-level data, provided on the ML-friendly Hugging Face platform. We present physics coverage and the generation, simulation, digitisation and reconstruction pipeline, describe format and access, and initial collider physics benchmarks.
\end{abstract}

\section{Introduction}

Particle physics is a field hungry for high quality simulation, to match the precision with which data is gathered at collider experiments such as the Large Hadron Collider (LHC). Comparing observed and predicted behaviour may hint at new phenomena, but may also be due to a range of sources of error in the simulation or data gathering process. As such, simulating the behaviour of both the expected physics processes and the detector itself (including possible miscalibrations) as truthfully as possible, is usually a key step in the discovery or exclusion of Beyond Standard Model (BSM) physics. Broadly, collider physics simulation falls into two classes: So-called ``full'' simulation or ``fast'' simulation. Both cases begin by predicting the probabilities of all immediate particle productions and decays, including effects like hadronisation. Thereafter, in full simulation, a physics engine steps through the interactions of each of these collision products with the material of the surrounding detector, calculating the effect of scattering, measurements, and further decays or production of particles. Subsequently, for the highest fidelity of data, this ``truth'' simulation must be digitised to produce a more realistic detector response, including the possibility of electronic noise, misalignments of equipment, limits on the range of energy that can be measured in a given time window, charge drifts, and so on. 

Full simulation is computationally onerous, and so in many cases it is possible to tolerate a loss of fidelity and produce a fast simulation. There are many promising ways to speed up particle simulation, such as using generative ML models, trained on full simulation. A popular approach is to parameterise the complex interactions of full simulation into a set of functions that can be applied to the collision products. An example may be to replace the probabilistic and noisy path of a charged particle through material in a magnetic field with a smeared helical trajectory. The helix can then be overlaid onto the detector, and various measurements or downstream interactions produced quickly. In this way, huge datasets of fast simulation data can be made available, enabling training large ML models on tens of millions to billions of samples. However, there may be a cost to using fast simulation: ML models are training on a parameterisation rather than realistic simulation. The impact of this bargain on the ability of so-trained models to detect subtle traces of new physics has been unclear. To that end, we propose a new full-simulation dataset, \colliderml{}, which aims to fill three roles:

\begin{enumerate}
    \item A repository for \textit{all} scales of physics objects, from generated partons, to energy deposits in material, through to reconstructed tracks and jets. These are realistically digitised, and may allow for studies of ML-aided algorithms for reconstruction of various stages in a collider workflow. In particular, we highlight multi-stage or multi-task models, anomaly detection models and foundation models as particularly suited to the \colliderml{} dataset.
    \item A playground for what is possible in a next-generation detector. \colliderml{} events are large and noisy, and detector components are borrowed from a range of proposed future technologies (e.g. high magnetic field, high-granularity calorimetry, pixel and strip inner tracker). To this end, we provide a variety of SM and BSM channels with fully-simulated (\textit{and non-repeated}) proton-collision pile-up, as well as a variety of miscalibrations.
    \item A way to study the impact of training models on fast simulation. In particular, are certain behaviours such as memorisation, generalisability and symmetry-preservation particular to these simplified datasets, or a general property of all physics simulation?
\end{enumerate}

\begin{figure}
    \centering
    % \hspace*{-0.25\linewidth}\includegraphics[width=1.5\linewidth]{img/release_schedule.png}
    \includegraphics[width=1.0\textwidth]{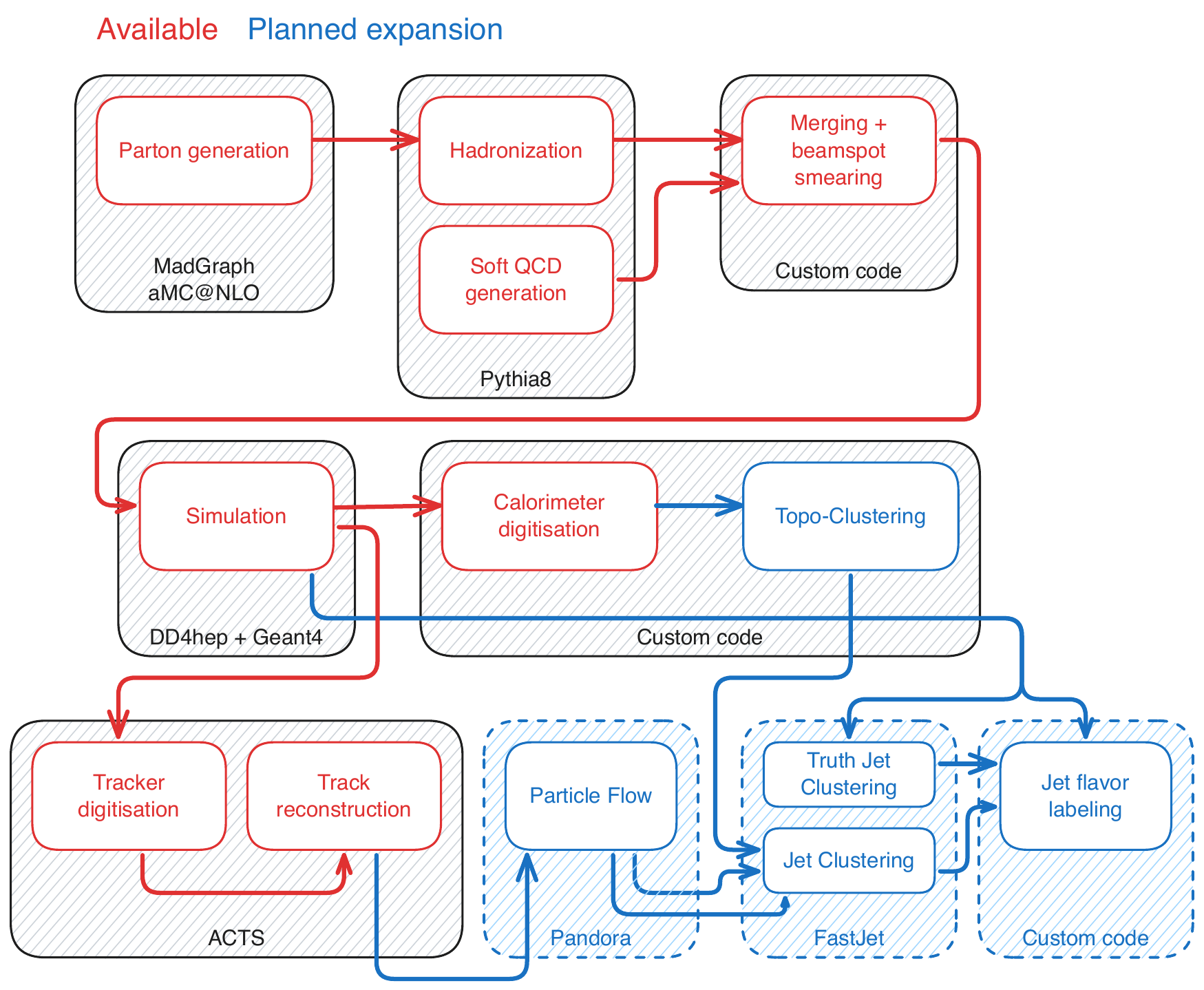}
    \caption{Technologies used to produce \colliderml{} and the objects in each release}
    \label{fig:release-schedule}
\end{figure}

We publicly release four sub-datasets, comprising \colliderml{}: over one million fully simulated, high-luminosity collision events, across ten production channels; ten million hard-scatter events across the same channels; ten million single-particle simulations; and three hundred thousand events that experience a range of miscalibrations and other challenging conditions. All events will eventually contain multiple scales in the simulation and reconstruction chain, from initial state particles to hits to analysis-level objects. In this note, we focus on the lower-level objects of the full pile-up production channels, which will be made available in \colliderml{} Release 1, with most available now on Hugging Face \cite{daniel_murnane_2025}. Release 2 and 3 are to contain the remaining reconstructed objects and miscalibrated conditions. This release schedule can be seen in \cref{fig:release-schedule}.

In this work, we detail how \colliderml{} was produced and the choices made in service of the above goals. We also present several metrics that should be useful in standardising the reporting of tracking tasks. In \cref{sec:simulation} we present the technology used for parton generation and simulation of particle trajectories and decays. In \cref{sec:digitisation}, we present the approach to digitisation of the inner detector and calorimeters. In \cref{sec:reconstruction}, the reconstruction of tracks is presented. Finally, in \cref{sec:data}, we sketch out the parquet file structure available to users, and how the data may be accessed.

\section{Previous Work}
\label{sec:previous-work}

Collider simulations are typically either \emph{full} (\geant{} \-based \cite{GEANT4:2002zbu,Allison:1035669} transport, possibly with realistic digitisation) or \emph{fast} (parameterised responses or learned surrogates). Fast approaches can scale to $10^7$–$10^8$ examples but replace stochastic transport with simplified functions (e.g. parameterised calorimeter responses). This speed–fidelity trade–off has enabled modern ML pretraining and benchmarking at scale, yet leaves open how much the simplifications bias learning and downstream inference. In particular, the more simplified the parameterisation, the less realistic fine-grained effects and objects become. Delphes \cite{Selvaggi:2014mya} is a popular approach to parameterised fast simulation. While recent developments include more sophisticated treatment of objects down to track-level (such as the covariance), realistic tracker and calorimeter energy deposits can not be produced in this way \cite{Selvaggi:2025delphes}.

\FloatBarrier
\begin{figure}
    \centering
    \hspace*{-1.5cm}
    \includegraphics[width=1.2\linewidth]{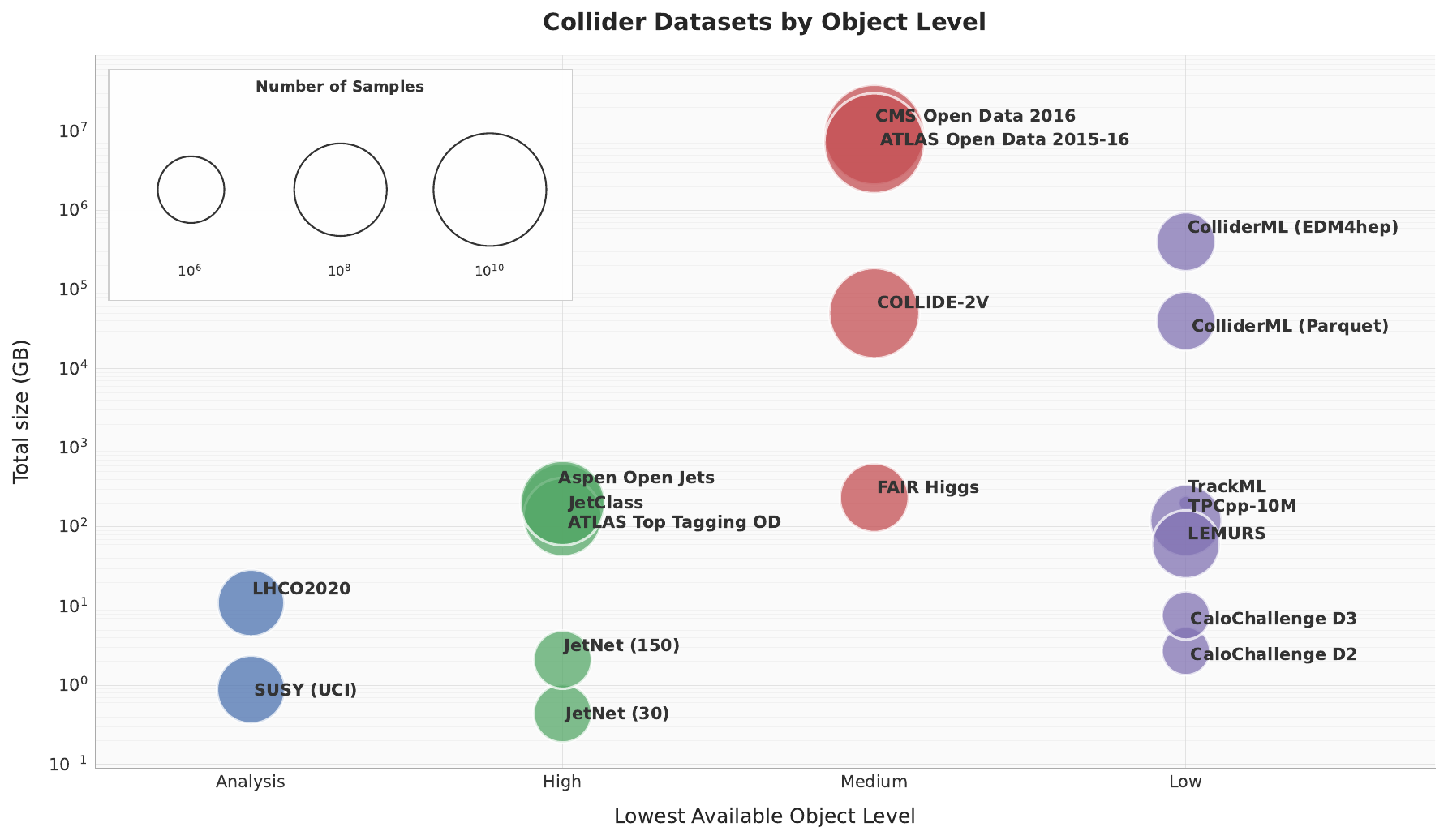}
    \caption{A comparison of \colliderml{} with others in the landscape of Collider Datasets. The values behind this visualisation are provided in \cref{app:datasets}.}
    \label{fig:collider_datasets}
\end{figure}

Beyond hand–tuned parameterisations, learned surrogates accelerate subdetectors with generative models (e.g.\ \emph{CaloGAN} for electromagnetic/hadronic showers \cite{Paganini_2018}; fast calorimeter variants within ATLAS/CMS fast-sim toolchains \cite{Aad_fastsim_2022,Bein_2024}). These methods trade exact transport for distributional fidelity and speed, often matching low-order moments but struggling with long tails, rare topologies, or high-granularity geometry effects—precisely the regimes where downstream ML is sensitive. In parallel, \emph{learned inference} frameworks use classifiers to correct simulation to data or unfold detector effects (e.g.\ \emph{OmniFold}’s iterative, simulation-aware reweighting \cite{Andreassen_2020}, related classifier-based likelihood-ratio estimation \cite{cranmer2016approximatinglikelihoodratioscalibrated,Brehmer_2018}). Together, these advances show how ML can both replace and repair simulation.

A rich ecosystem of collider physics datasets exists at \emph{analysis} (event-level statistics), \emph{high} (particle flow objects), \emph{medium} (tracks and calorimeter clusters), and \emph{low} (detector hits/voxels) abstraction levels. We summarise representative datasets and sizes in \cref{fig:collider_datasets}. At high scale, JetNet and JetClass \cite{jetclass} provide $10^6$–$10^8$ jet samples for supervision and generative modelling; ATLAS Top Tagging Open Data extends supervised benchmarks; and SUSY (UCI) offers classic tabular high–level features. Low–level resources include TrackML \cite{trackml_acc,trackml_through} (silicon tracking) and the CaloChallenge electromagnetic/hadronic showers~\cite{calochallenge2022}.

\paragraph{Why \colliderml{}}
Most public sets either expose only high/medium abstractions or isolate a single subdetector in controlled conditions; few provide full-simulation, full-detector events with realistic, non–repeated pile–up, miscalibrations, and all intermediate objects in one place. The scale of fast–simulation datasets has also encouraged training very large models where over-fitting may emerge below $\mathcal{O}(10^6)$ training examples, muddying generalisation claims. \colliderml{}\ targets these gaps with a multi–scale, HL–LHC–like environment and controllable realism knobs to study memorisation, inductive biases, and symmetry preservation under full simulation.

In particular, we expect ColliderML to be well-suited to the testing of foundation model generalisability between fine-grained and high-level objects, where previous models focussed only on low-level \cite{huang2024languagemodelparticletracking} or high-level \cite{bhimji2025omnilearnedfoundationmodelframework, Birk_2024, golling2024maskedparticlemodelingsets}. It should be sufficiently realistic to study the scaling of ML-based algorithms for tracking and clustering \cite{Duarte_2022, DeZoort_2021, gnn_sylvain, influencer, liu2023hierarchicalgraphneuralnetworks, Ju_2021, Burleson:2882507, vanstroud2024transformerschargedparticletrack, Kieseler_2020}, fast simulation \cite{kansal2022particlecloudgenerationmessage,krause2023caloflowiifasteraccurate,Buhmann_2023} and particle flow \cite{kakati2025hgpflowextendinghypergraphparticle,mokhtar2025machinelearningbasedparticleflowalgorithm} in conditions very similar to those foreseen in Phase 2 upgrades of LHC detectors. Finally, it may serve as a way to examine physics-motivated models which have historically focussed on testing in fast simulation \cite{Gong_2022,Bogatskiy_2024,spinner2024lorentzequivariantgeometricalgebratransformers,Murnane_2023,murnane2023equivariantgraphneuralnetworks,thais2023equivarianceneedcharacterizingutility}.

Relative to prior work, \colliderml{}\ aims to make available (i) joint availability of all object levels per event for cross–stage studies; (ii) realistic pile–up and detector effects (future releases will also include miscalibrations) suitable for stress–testing reconstruction and ML; and (iii) a release plan that brings lower–level objects first, followed by higher–level reconstructions, to enable community baselines and foundation-model pretraining.

\section{Simulation}
\label{sec:simulation}

\subsection{Included Physics}

In the first release of \colliderml{} (high pileup), we make available ten production channels, listed in \cref{tab:production_channels}, with brief notes for why each channel was chosen. In particular, we include those Standard Model channels that are useful for precision studies of couplings and masses in the electroweak sector, Higgs sector, and top sector. We also include a jet-heavy pile-up channel for performance studies of reconstruction algorithms. For BSM physics, we select three distinct signatures that map to key machine learning challenges in reconstruction and anomaly detection. First, we include a gauge-mediated supersymmetry breaking model (GMSB) \cite{Dimopoulos_1997} featuring a long-lived neutralino next-to-lightest supersymmetric partner (NLSP) decaying to a gravitino dark matter candidate (with a lifetime of $c\tau \approx 100$ mm). This provides a displaced vertex and missing energy signature, serving as a target for new physics searches and non-standard tracking, such as in the Run 3 search for displaced leptons with the ATLAS detector \cite{Aad_2025}.

Second, a Hidden Valley (Dark QCD) channel is included \cite{Strassler_2007}, modelled as a $Z'$ mediator decaying to a dark sector with admixed stable and unstable dark hadrons. This produces semi-visible jets—objects with localised missing energy and complex substructure, serving for example as a benchmark for anomaly detection \cite{cmsvisiblejets2022}. Finally, the well-motivated Z-prime heavy resonance model \cite{Langacker_2009} is included to benchmark standard supervised bump-hunting and resolution studies. These exotic signatures are central to the physics programmes of proposed future facilities such as the EIC and FCC \cite{Abdul_Khalek_2022, Mangano:2651294, https://doi.org/10.17181/cern.35ch.2o2p}.

\begin{table}[!h]
    \renewcommand{\arraystretch}{1.7} % Adjust row spacing
    \setlength{\tabcolsep}{10pt} % Adjust column spacing
    \hspace*{-1cm}\begin{tabular}{r|l|l|m{8cm}}
        \hline
        \textbf{ID} & \textbf{Process Label} & \textbf{Process Type} & \textbf{Comments} \\
        \hline
        1 & \texttt{ttbar}      & $pp \to t\bar{t}$  & Top quark production; important for SM tests and complex final states. \\
        2 & \texttt{zee}   & $pp \to Z \to e^+e^-$  & Drell-Yan process; detector calibration and electroweak probe. \\
        3 & \texttt{zmumu}   & $pp \to Z \to \mu^+\mu^-$  & Drell-Yan process; detector calibration and tracking probe. \\
        4 & \texttt{dihiggs}    & $gg \to HH$  & Gluon fusion di-Higgs; benchmark for high-Luminosity studies. \\
        5 & \texttt{diphoton}    & $pp \to \gamma\gamma$   & Diphoton production; loop-induced QCD/EW process. \\
        6 & \texttt{multijet}     & $pp \to \mathrm{jets}$  & QCD production; dominant background for hadronic signatures. \\
        7 & \texttt{ggf}        & $pp \to H$  & Gluon-gluon fusion; main Higgs production channel. \\
        8 & \texttt{susy}       & $pp \to \tilde{g} \tilde{g}$  & GMSB gluino production; $\tilde{\chi}^0_1$ NLSP provides displaced vertex benchmark. \\
        9 & \texttt{zprime}     & $pp \to Z'_{SSM}$  & Heavy resonance; standard candle for bump hunting. \\
        10 & \texttt{hiddenvalley} & $pp \to Z' \to \text{dark}$ & Dark QCD sector; semi-visible jets for anomaly detection. \\
        \hline
    \end{tabular}\vspace{5pt}
    \caption{Listing of production channels in \colliderml{} (high pileup). BSM channels (8-10) are selected to represent distinct reconstruction challenges.}
    \label{tab:production_channels}
\end{table}

Also included are hard-scatter-only samples (i.e. with no pile-up) and particle gun channels, where single particles are fired at set energy levels in a uniformly sampled initial direction. In these \texttt{singleparticle} channels, we provide single $\mu^\pm$, single $e^\pm$, single $\pi^\pm$, single $K^\pm$ and single $\gamma$, across several discrete and continuous energy regimes.

Hard-scatter matrix element calculation is state-of-the-art, with next-to-leading order (NLO) calculation of the main process and the first two jets, in Madgraph aMC@NLO v3.58 \cite{Alwall2014MG5aMC}. Loop-induced processes (GGF and di-Higgs) are modelled at LO in Madgraph. Showering is provided with \pythia\texttt{.313} \cite{Sjostrand2015Pythia8}, with FxFx matching for NLO processes, and CKKW matching for LO processes. BSM channels are natively modelled and showered at LO within \pythia.

We are cautious to highlight specifically the case of vector boson fusion and scattering, which is not well-modelled in the default setup of aMC@NLO$\rightarrow$ \pythia{} pipeline used above \cite{J_ger_2020, höche2021studyqcdradiationvbf}. As such, we have avoided channels involving these phenomena, for example by excluding $VV\rightarrow HH$, allowing only $gg\rightarrow HH$ as the di-Higgs production channel\footnote{We note that VBS and VBF channels are important to include, and will release these in future, making use of POWHEG-BOX for parton generation and Pythia's dipole recoil scheme when showering. This has been shown to agree much more closely with measurement \cite{höche2021studyqcdradiationvbf}.}.

\paragraph{Pile-up generation.}
In the Phase 2 upgrades of the ATLAS \cite{CERN-LHCC-2017-021,ITkStripTDR} and CMS \cite{CERN-LHCC-2017-009} experiments, for each bunch crossing there will be an expected 200 proton-proton collisions. Most of these will be soft QCD scatters, and not of interest to most physics analyses. However, they must be modelled and simulated, in order for the hard-scatter collision of interest to be distinguished from this background.  The typical workflow adopted by collaborations is to produce a large pool $P$ of pile-up simulations, sample $N \ll P$ so-called pile-up ``sub-events'', and combine with a hard-scatter sub-event to produce a full pile-up event. This is computationally efficient, but can lead to significant overlap between event content. This may not be an important effect in datasets that only consider analysis-level objects, but at low-level, it would lead to the same tracker hits and calorimeter contributions appearing tens or hundreds of times across a dataset. While it remains to be exhaustively demonstrated that this repetition can be memorised, we are cautious to avoid this possibility in \colliderml.

\begin{figure}
    \centering
    \includegraphics[width=0.7\linewidth]{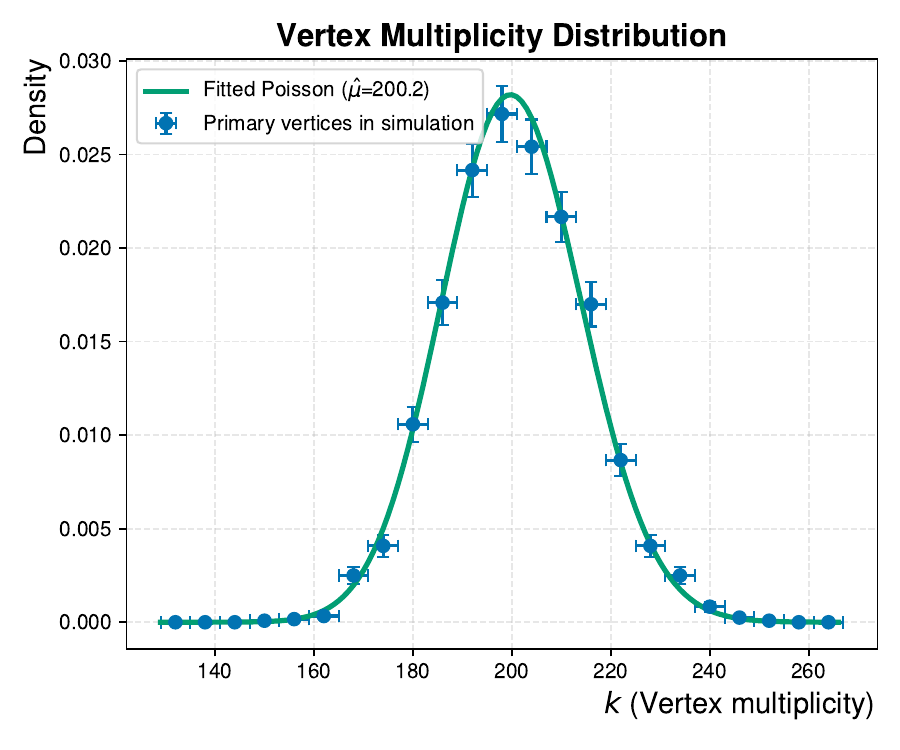}
    \caption{Number of primary vertices (single hard-scatter and N pile-up sub-events) across 2000 events. A poisson curve fitted to the counts validates the events follow the expected distribution.}
    \label{fig:pileup_poisson}
\end{figure}

As a result, every pile-up sub-event is generated independently (giving a pool of approximately 200 million parton sub-events), and combined without repetition \textit{before} simulation. As pile-up is unlikely to be the focus of detailed analysis, we use \pythia{}'s native leading-order matrix element calculation to generate the process and subsequently shower within \pythia{}. For each hard-scatter process, $N$ pile-up events are sampled from a poisson distribution with $\lambda=200$ and merged into a single \texttt{hepmc} file. This can be seen to behave as expected in \cref{fig:pileup_poisson}.

\paragraph{Vertex smearing}
As part of the sub-event merging of showered stable particles (i.e. one hard-scatter and N pile-up), we simulate the effect of a finite beam spot. Each sub-event is shifted randomly according to a Gaussian distribution centred at the origin. The spatial and temporal widths are set to $\sigma_{x,y} = 12.5\;\mu$m, $\sigma_z=55.5$\;mm and $\sigma_t = 185$\;ps.

\subsection{Detector Description}
\label{sec:odd}

The detector geometry used is that of the OpenDataDetector (ODD) \cite{ODD2023TrackingSystem}, a project driven by the \acts{} \cite{acts} collaboration, as an experiment-agnostic way to provide realistic physics simulations in high-luminosity and future collider environments. The ODD combines design choices from existing and future detectors: a wide-coverage silicon inner tracker (IT) inspired by the ATLAS Phase 2 ITk \cite{ATLAS_ITk_Pixel_TDR_2017, ATLAS_ITk_Strip_TDR_2017}, an electromagnetic calorimeter (ECal) resembling several FCCee detector proposals and a hadronic calorimeter (HCal) similar to the CMS HGCal subdetector and future detectors CLD, AHCAL and SiD  \cite{CLD_Note_2019, CMS_HGCAL_TDR_2017, Bacchetta2019CLD, Adloff2010CALICEAHCAL, SiD_LOI_2009}. A muon detection system is planned for inclusion in future. The inner tracker consists of two subsystems: high-precision pixel silicon in the innermost layers, and lower-precision strip silicon in the outermost layers. Together, the tracker detector layers provide at least 12 layers of sensitive material across a pseudorapidity of $\eta \in [-3, 3]$, and 8 layers across $\eta \in [-3.5, 3.5]$.

\begin{figure}
    \centering
    \begin{subfigure}[b]{1.0\textwidth}
    \hspace*{-1.5cm}% <-- tweak this
    \includegraphics[width=1.2\textwidth]{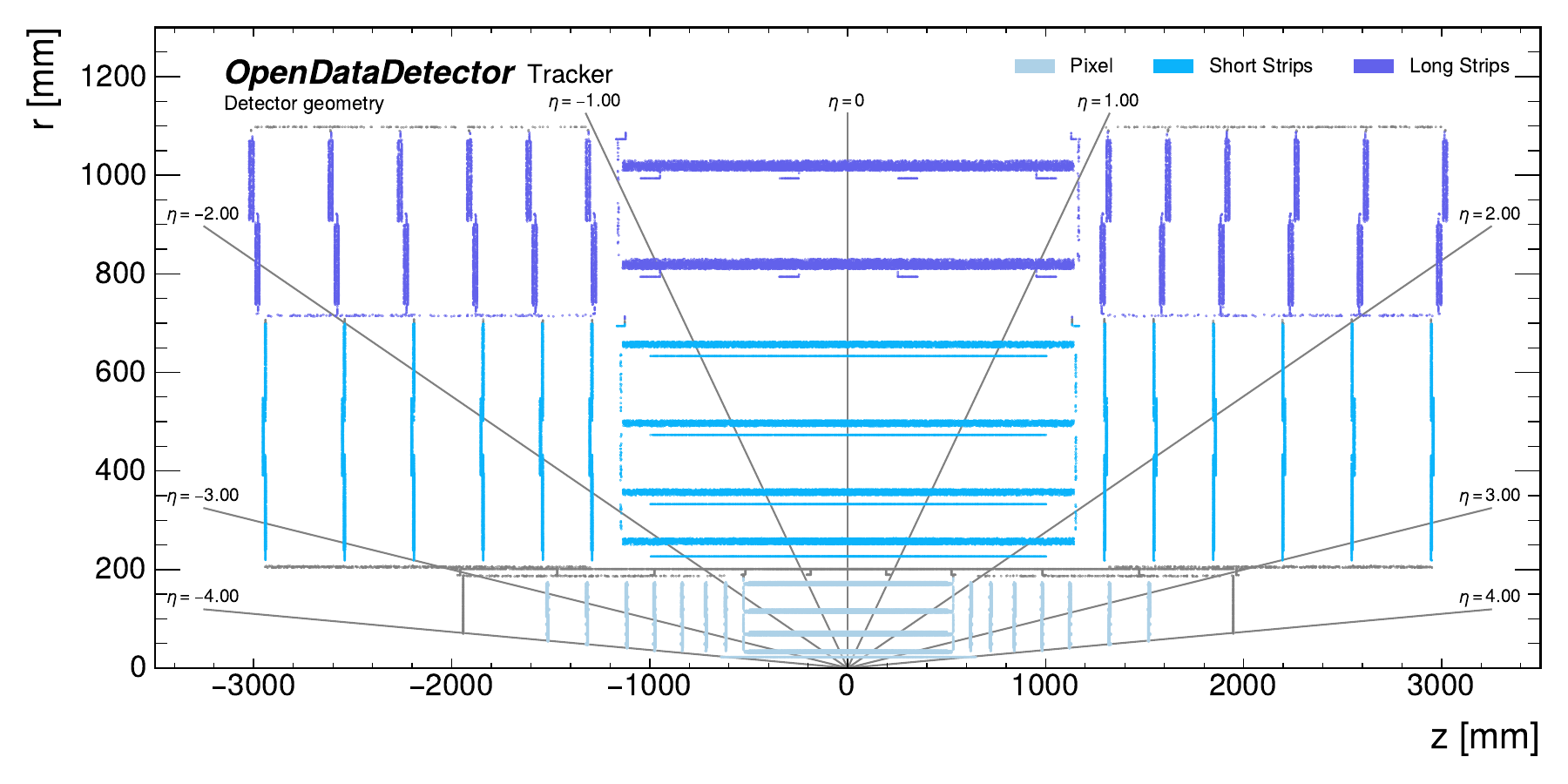}
    \caption{Longitudinal view of the ODD tracking system}
    \end{subfigure}
    
    \begin{subfigure}[b]{1.0\textwidth}
    \hspace*{-1.5cm}% <-- tweak this
    \includegraphics[width=1.2\textwidth]{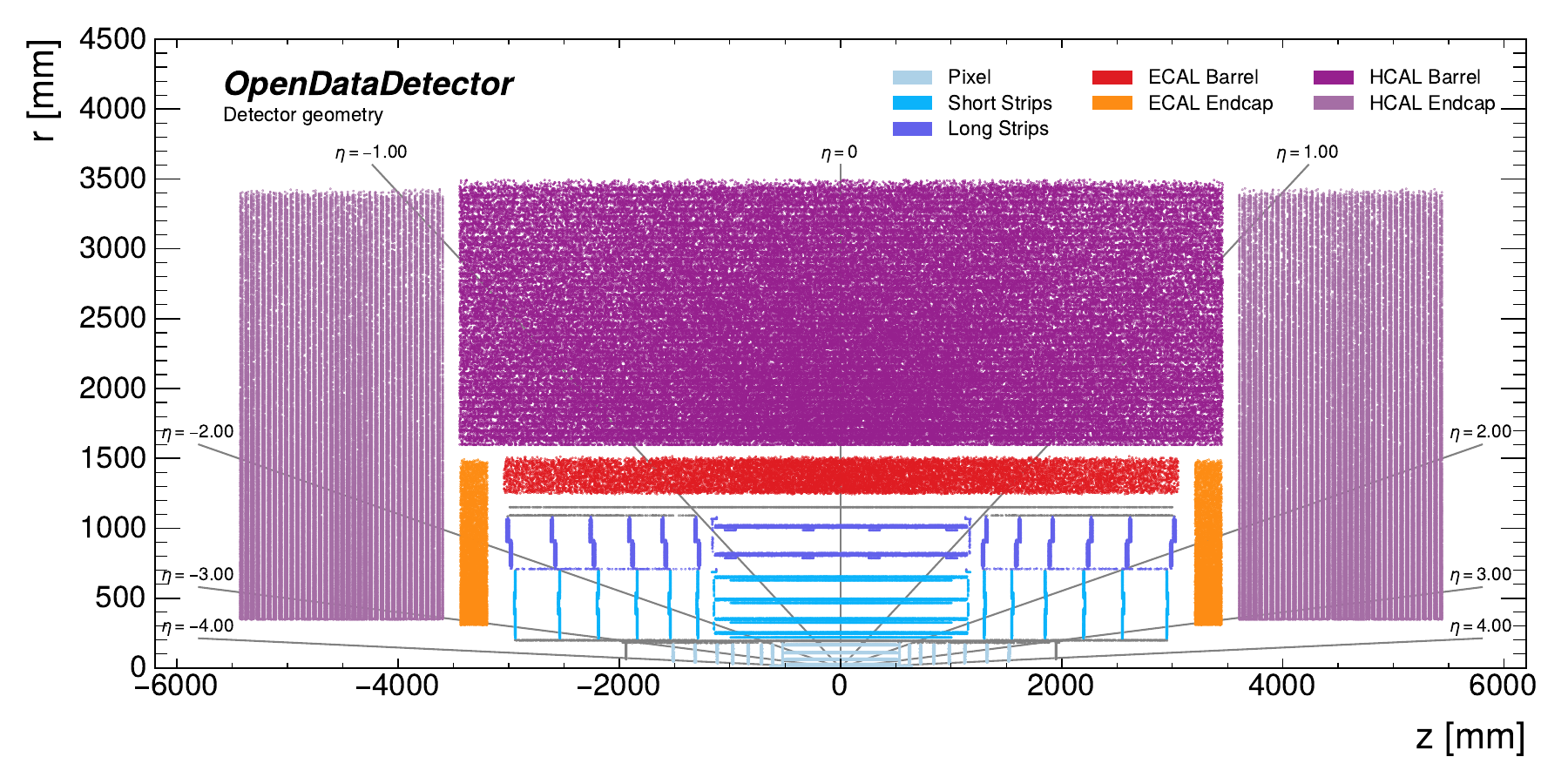}
    \caption{Longitudinal view of the ODD including the calorimeters}
    \end{subfigure}
    
\end{figure}

\begin{figure}
    \centering
    \begin{subfigure}[b]{1.0\textwidth}
        \centering
        \includegraphics[width=0.7\textwidth]{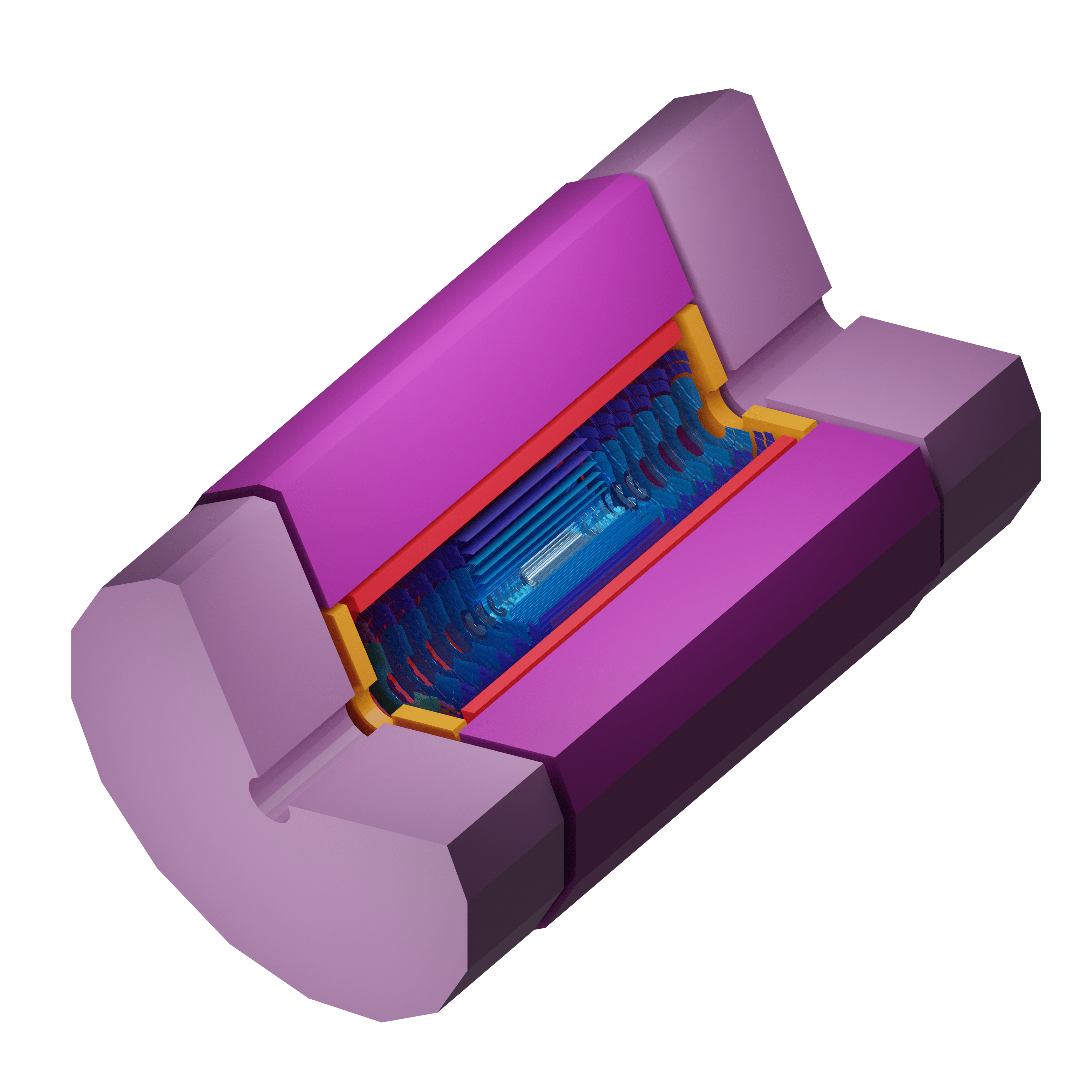}
        \caption{Isometric 3D view of the ODD tracking and calorimeter systems}
        \label{fig:tracker-coverage}
    \end{subfigure}
    \caption{The ODD inner tracker is composed of highly granular silicon pixels in the innermost layers, short silicon ``strixel'' strips in middle layers, in longer 2D resolution strips in the outermost layers. The tracker is expected to deliver well-reconstructed tracks up to $\eta \leq 3.5$ \cite{gessinger2023open}. The calorimeters are highly granular, with at least 27\;$X_0$ depth for ECal, and 10\;$\lambda_0$, for barrel and the endcaps.}
\end{figure}

The electromagnetic calorimeter comprises 48 layers of 0.5\;mm thick silicon sensors, each divided into cells of size 5.1\;mm $\times$ 5.1\;mm and
interlaid with 1.9\;mm tungsten absorption layers. The readout, spacers, and addition material accounts for additional thickness of 2.65\;mm per layer. The hadronic calorimeter is composed of 36 layers of 3\;mm thick polystyrene scintillator, divided into cells of 30 mm $\times$ 30\;mm size. The
absorber is 30\;mm thick steel and the remaining 16\;mm corresponds to the readout.

This work uses the default OpenDataDetector solenoid field setting, which has
been slightly updated from previous versions of the detector (see
\cite{gessinger2023open}). The field is now generated using a solenoid producing
3\;T at the center of detector, and is configured with an ambient field outside
of the solenoid of 0.5\;T. The value of 3\;T was chosen to present a novel
working point in the track curvature, interpolated between the 2 T magnetic
field of the ATLAS detector \cite{CERN-LHCC-97-021} and the 4 T of the CMS detector \cite{CERN-LHCC-97-010}.
The outer boundary of the sensitive tracking material is taken to be
$r \leq 1080$\;mm and $z \in
[-3030\textnormal{\;mm}, 3030\textnormal{\;mm}]$. This is used to define the
truth handling of particles incident on the calorimeter, described in
\cref{sec:geant}.

\subsection{Particle Propagation and Interactions}
\label{sec:geant}

To simulate the path taken by each primary particle produced by \pythia{}, and the decays to secondary particles, \geant{} \cite{AGOSTINELLI2003250} is used. In particular, we make use of the \ddforhep{} library \cite{Frank2014DD4hep} provided as part of the \keyforhep{} framework, which includes convenience functions to consume the detector geometry description, as well as to specify the handling of truth, and finally provides the standard \edmforhep{} data model \cite{Gaede:2869508} to capture all information about the event simulation.

Some additions were made to the standard \ddforhep{} configuration to better suit the requirements of the \colliderml{} dataset\footnote{We expect these changes to be available in the main branch of \ddforhep{} and a version of the LHC Computing Grid (LCG) software release in the near future.}. Multithreading is implemented in an experimental mode, which increases the possible throughput of simulation by around 4x per CPU node. This is due to the simulation being memory-constrained in multi-process mode. Simulation is the most expensive stage of the \colliderml{} pipeline, and this development increases the number of available events from around 300,000 to over 1,000,000, with the available compute budget. An examination of the scaling behaviour is provided in \cref{app:dd4hep_mt}.

The second customisation is in order to allow for more particle decay truth to be retained. In default \ddforhep{} simulation, the vast majority of particles are discarded, leading to all subsequent daughter particle hits being attributed to a small set of parents. For \colliderml{}, we introduce two extra levels of truth handling that are more conservative, tending to include all particle details where there could be any possible use case for it, or where discarding may cause doubt or confusion. A full description of the default and custom handling is given in \cref{app:truth_handling}. In short, the particles retained at simulation level are provided according to the flowchart in \cref{fig:dd4hep_full_truth_handling}. In that case, only ``book-keeping'' particles created during showering and simulation, or very low-energy particles that are produced and decay in the calorimeter are discarded. Any particle that leaves an energy deposit in the sensitive material is retained in full.

\begin{figure}
    \centering
    \includegraphics[width=0.8\linewidth]{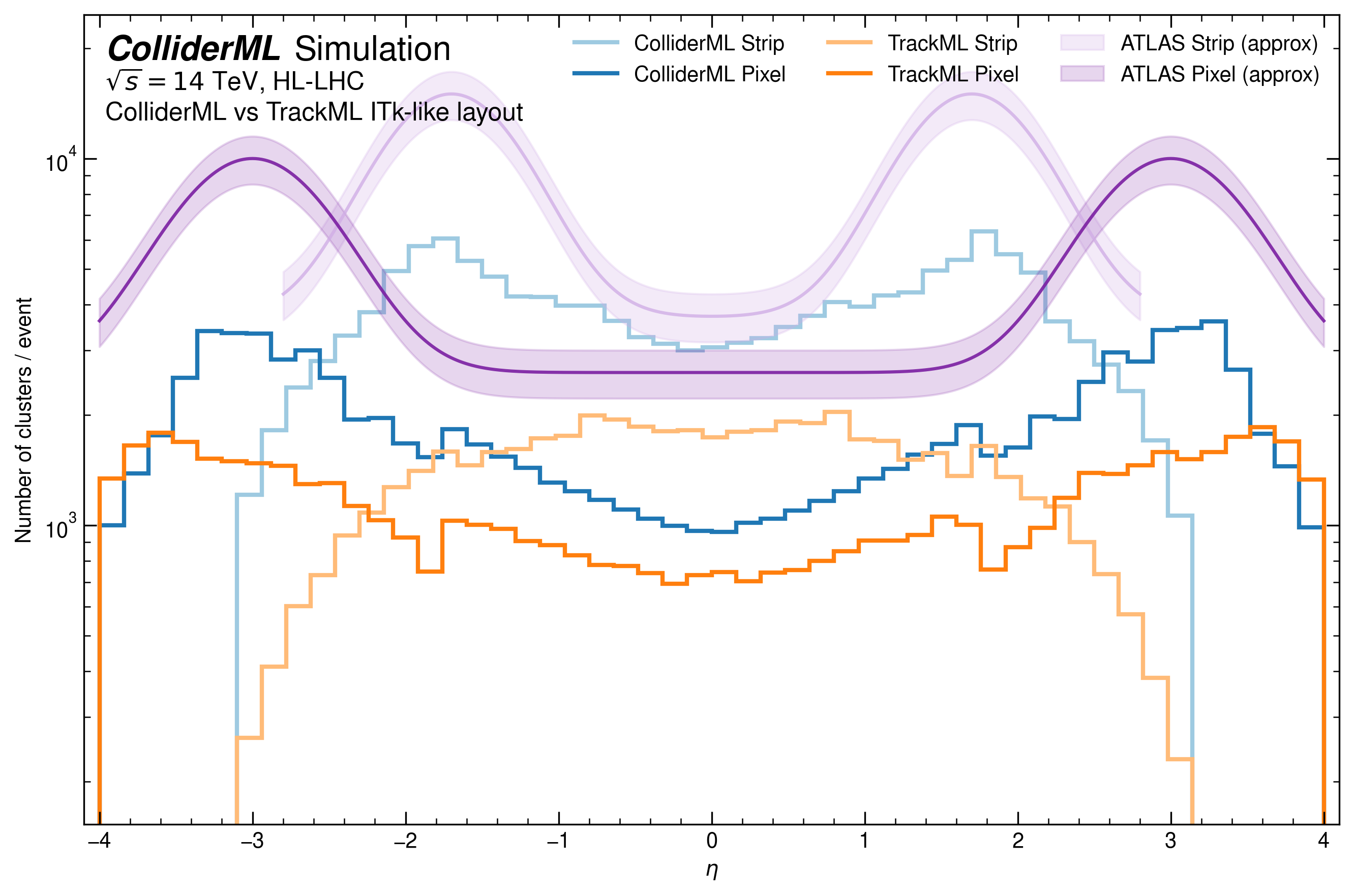}
    \caption{The distribution of inner tracker hits vs pseudorapidity for the simulated TrackML dataset (orange), ATLAS Phase 2 ITk (purple), and \colliderml{} (blue). ATLAS statistics derived from \cite{Aad_2025_tracking}}
    \label{fig:eta_distribution_hit_number_comparison}
\end{figure}

\begin{figure}
    \centering
    \includegraphics[width=0.9\linewidth]{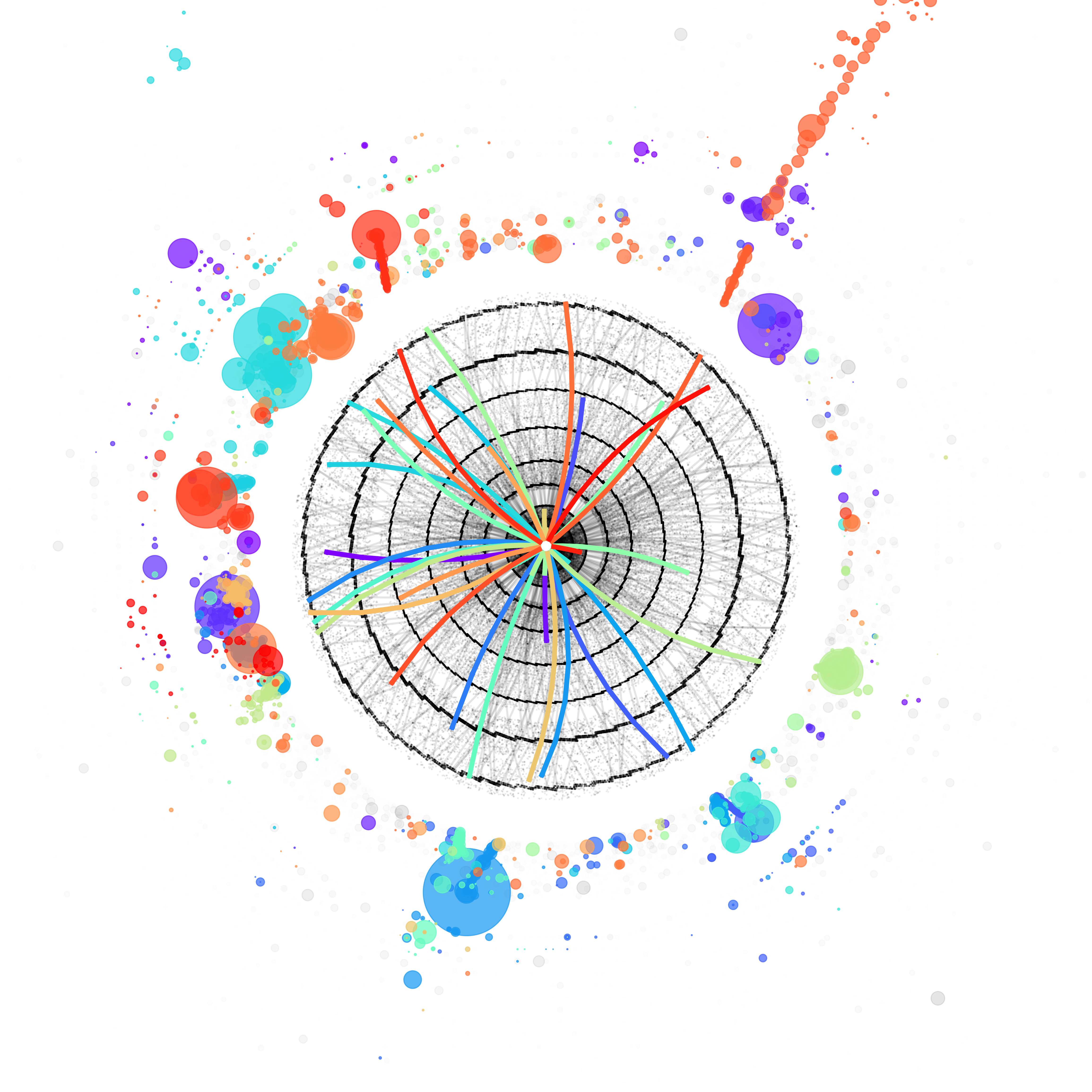}
    \label{fig:tracker_and_calo_event_viz}
    \caption{A $t\bar{t}$ event display of \geant{} simulation of particle trajectories through the ODD. In grey, the full collection of energy deposits left across the ODD. A sample of particles are selected, and their IT and calorimeter deposits coloured accordingly, with each calorimeter contribution energy by the size of the circle.}
\end{figure}

\section{Digitisation}
\label{sec:digitisation}

The \colliderml{} dataset provides highly realistic digitisation of to imitate the detector response in the inner tracker and both calorimeters.

\subsection{Tracker Digitisation}
\label{sec:tracker_digitisation}

\begin{figure}
    \centering
      \tikzsetnextfilename{clustering}

  \begin{tikzpicture}[scale=1.0,line width=0.75pt]

  \def\yl{0.8cm}
  \def\zl{0.9cm}
  \def\x{({cos(0)*1cm},{sin(0)*1cm})}
  \def\y{({cos(45)*\yl},{sin(45)*\yl})}
  \def\z{({cos(90)*\zl}, {sin(90)*\zl})}

  \def\w{3.5}
  \def\h{3}
  \def\d{1}
  
    \def\nx{5}
    \def\ny{5}
    \pgfmathsetmacro{\dx}{\w/(\nx+1)}
    \pgfmathsetmacro{\dy}{\h/(\ny+1)}
    \pgfmathsetmacro{\nxt}{int(\nx+1)}
    \pgfmathsetmacro{\nyt}{int(\ny+1)}

  \begin{scope}[scale=1,
                      x={(\x)},y={(\y)},z={(\z)}
                      ]
    \begin{scope}[shift={(-0.5,-0.5,-0.5)}]
      \draw[->] (0,0,0) -- (1,0,0) node[at end,right] {$x$};
      \draw[->] (0,0,0) -- (0,0.5,0) node[at end,above right] {$y$};
      \draw[->] (0,0,0) -- (0,0,1) node[at end,above] {$z$};
    \end{scope}

    \begin{scope}[canvas is xz plane at y=\h]
      \draw (0,0) rectangle (\w, \d);
    \end{scope}

    \begin{scope}[canvas is yz plane at x=0]
      \draw (0,0) rectangle (\h,\d);
    \end{scope}

    \begin{scope}[canvas is xy plane at z=\d]
      \coordinate (bl) at (0,0);
      \coordinate (tr) at (\w,\h);

      \fill[white,opacity=0.8] (bl) rectangle (tr);
      
      \coordinate (x0) at (bl);
      \coordinate (x\nxt) at (bl-|tr);

      \coordinate (y0) at (bl);
      \coordinate (y\nyt) at (bl|-tr);

      \foreach \i in {1,...,\nx} {
        \coordinate (x\i) at ($(bl) + ({\i*\dx},0)$);
      }

      \foreach \i in {1,...,\ny} {
        \coordinate (y\i) at ($(bl) + (0,{\i*\dy})$);
      }

      \coordinate (ixi0) at ({0.13*\w},{0.35*\h});
      \coordinate (ix) at ($(y3-|x5)+(0.1,0.1)$);
      % \fill[Purple] (ix) circle(2pt);
       
      \begin{scope}[/pgfplots/colormap/viridis,
            cmap/.style={/utils/exec={
                    \pgfplotscolormapdefinemappedcolor{#1}},
                fill=mapped color}]
              
          \fill[cmap=250] (x1|-y2) rectangle (x2|-y3);
          \fill[cmap=500] (x2|-y2) rectangle (x3|-y3);
          \fill[cmap=750] (x3|-y2) rectangle (x4|-y3);
          \fill[cmap=1000] (x4|-y3) rectangle (x5|-y4);
      \end{scope}

      \draw (bl) rectangle (tr);
      
      \begin{scope}[line width=0.5pt]
        \foreach \i in {1,...,\nx} {
          \draw (x\i) -- (x\i|-tr);
        }

        \foreach \i in {1,...,\ny} {
          \draw (y\i) -- (y\i-|tr);
        }
      \end{scope}

      \begin{scope}[Red,line width=1pt]
        \draw[] (x0|-y2) rectangle (x1|-y3);
        \draw[] (x4|-y2) rectangle (x5|-y3);
        \draw[] (x5|-y3) rectangle (x6|-y4);
      \end{scope}

      \draw[line width=1pt,dashed] (ixi0) -- (ix);

    \end{scope}

    \begin{scope}[canvas is xy plane at z=0]
      \coordinate (ixi) at ({0.13*\w},{0.35*\h});
    \end{scope}

    \begin{scope}[canvas is xz plane at y=0]
      \draw[fill=white,fill opacity=0.3] (0,0) rectangle (\w, \d);
    \end{scope}

    \begin{scope}[canvas is yz plane at x=\w]
      \draw (0,0) rectangle (\h,\d);
    \end{scope}

    % \node (_pdir) at (1.6,0.3,0.7) {};
    \coordinate (pdir) at ($(ix) - (ixi)$);

    \begin{pgfonlayer}{background}
      \draw[line width=1pt] (ixi) -- ($(ixi)-0.25*(pdir)$);
      \begin{scope}[canvas is xy plane at z=0]
        \fill[white!90!black,opacity=0.7] (0,0) rectangle (\w,\h);
        % \fill[LimeGreen] (ixi) circle(2pt);
      \end{scope}
      % \draw[line width=1pt] ($(ix)-2*(pdir)$) -- (ix);
      \draw[line width=1pt] (ixi) -- (ix);
    \end{pgfonlayer}

    \draw[->,line width=1pt] (ix) -- ($(ix)+0.6*(pdir)$) node[at end,right] {track};

  \end{scope}

  \begin{scope}[scale=1,
                      shift={(7,0)}
                      ]
    \begin{scope}[shift={(-0.75,-0.75)}]
      \draw[->] (0,0) -- (1,0) node[at end,right] {$x$};
      \draw[->] (0,0) -- (0,1) node[at end,above] {$y$};
    \end{scope}

    \coordinate (bl) at (0,0);
    \coordinate (tr) at (\w,\h);

    \fill[white,opacity=0.8] (bl) rectangle (tr);
    
    \coordinate (x0) at (bl);
    \coordinate (x\nxt) at (bl-|tr);

    \coordinate (y0) at (bl);
    \coordinate (y\nyt) at (bl|-tr);

    \foreach \i in {1,...,\nx} {
      \coordinate (x\i) at ($(bl) + ({\i*\dx},0)$);
    }

    \foreach \i in {1,...,\ny} {
      \coordinate (y\i) at ($(bl) + (0,{\i*\dy})$);
    }

    \coordinate (ix) at ($(y3-|x5)+(0.1,0.1)$);
      
    \begin{scope}[/pgfplots/colormap/viridis,
          cmap/.style={/utils/exec={
                  \pgfplotscolormapdefinemappedcolor{#1}},
              fill=mapped color}]
            
        \fill[cmap=250] (x1|-y2) rectangle (x2|-y3);
        \fill[cmap=500] (x2|-y2) rectangle (x3|-y3);
        \fill[cmap=750] (x3|-y2) rectangle (x4|-y3);
        \fill[cmap=1000] (x4|-y3) rectangle (x5|-y4);
    \end{scope}

    \draw (bl) rectangle (tr);
    
    \begin{scope}[line width=0.5pt]
      \foreach \i in {1,...,\nx} {
        \draw (x\i) -- (x\i|-tr);
      }

      \foreach \i in {1,...,\ny} {
        \draw (y\i) -- (y\i-|tr);
      }
    \end{scope}

    \begin{scope}[Red,line width=1pt]
      \draw[] (x0|-y2) rectangle (x1|-y3);
      \draw[] (x4|-y2) rectangle (x5|-y3);
      \draw[] (x5|-y3) rectangle (x6|-y4);
    \end{scope}

    \coordinate (ixi) at ({0.13*\w},{0.35*\h});
    \coordinate (pdir) at ($(ix) - (ixi)$);

    \draw[line width=1pt] ($(ixi)-0.4*(pdir)$) -- (ixi);
    \draw[dashed,line width=1pt] (ixi) -- (ix);
    \draw[->,line width=1pt] (ix) -- ($(ix)+0.5*(pdir)$) node[at end,right] {track};

  \end{scope}

  % \draw []
  \node[draw,rectangle,inner sep=0,minimum width=0.3cm,minimum height=0.2cm,line width=1pt,Red] 
  (rrect) at(4,3.5) {};
  \node[right=0mm of rrect] {energy below threshold};

  \end{tikzpicture}
    \caption{Illustration of how a simulated charged particle intersection can
    be \emph{digitised} by discretising the deposited energy into a fixed
    segmentation on a sensor.}
    \label{fig:digitisation}
\end{figure}
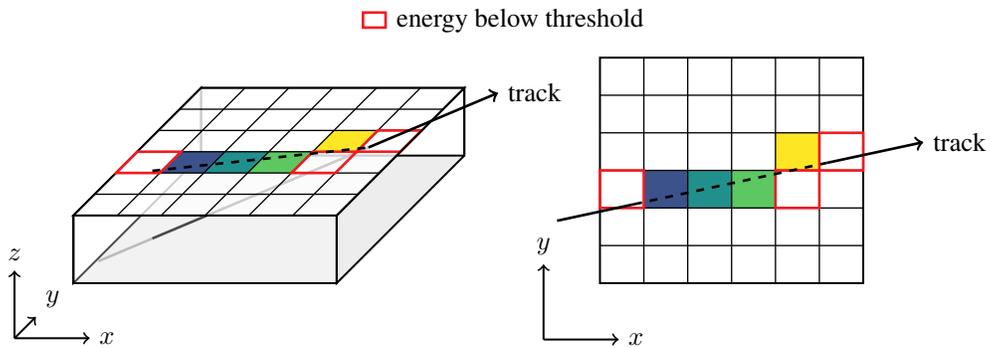

Hits in the tracker of the ODD are digitised with a geometric channelisation
approach using \acts{}. In this process, the position and direction of each particle
are extracted from the hits that were created during simulation. The particle
path is then projected onto the the flat sensor surfaces. 

\begin{figure}
    \centering
    \begin{subfigure}[b]{0.45\textwidth}
    \centering
      \tikzsetnextfilename{strip_digi}

  \begin{tikzpicture}[scale=0.6]

    \begin{scope}[shift={(10,0)}]

      \fill[Orange] (-1,0) rectangle (5,3);
      \fill[fill=Cyan,line width=2pt] (0,0) --++(4,0) --++(1,3) --++(-6,0) -- cycle;

      \foreach \i in {0,...,30}{
        \draw[line width=1pt] (\i*0.2-1,0) --++(0,3);
      }

      \foreach \j in {0,...,3}{
        \draw[line width=1pt] (-1,\j*1) --++(6,0);
      }

      \node[Cyan, anchor=south] at (2,3) {keep};
      \node[Orange,anchor=north east] at (5,0) {discard};
    \end{scope}

  \end{tikzpicture}
    \caption{Short strip digitisation}
    \label{fig:strip_digi_ss}
    \end{subfigure}
    \begin{subfigure}[b]{0.45\textwidth}
    \centering
      \tikzsetnextfilename{strip_digi}

  \begin{tikzpicture}[scale=0.6]

    \begin{scope}[shift={(0,0)}]
      \fill[Orange] (-1,0) rectangle (5,3);
      \fill[fill=Cyan,line width=2pt] (0,0) --++(4,0) --++(1,3) --++(-6,0) -- cycle;

      \foreach \i in {0,...,30}{
        \draw[line width=1pt] (\i*0.2-1,0) --++(0,3);
      }

      \node[Cyan, anchor=south] at (2,3) {keep};
      \node[Orange,anchor=north east] at (5,0) {discard};
    \end{scope}

  \end{tikzpicture}
    \caption{Long strip digitisation}
    \label{fig:strip_digi_ls}
    \end{subfigure}
    \caption{Illustration of the approach used to digitise the endcap sensors in the strip detectors using a rectangular segmentation.}
    \label{fig:strip_digi}
\end{figure}
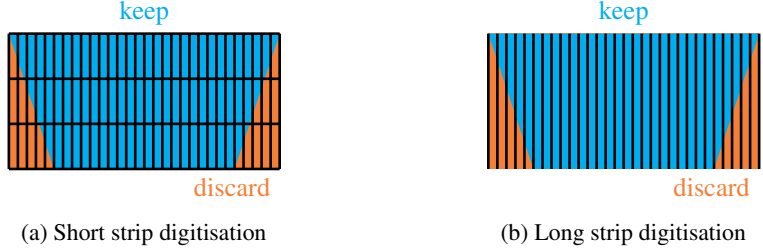

A segmentation is used to mimic the segmented readout found in a physical
particle sensor. For \colliderml{}, only rectangular segmentations are used for
both the rectangular barrel as well as the trapezoidal strip sensors found in
the endcap. An illustration of this rectanuglar segmentation is applied is found
in \autoref{fig:strip_digi}. The numerical values of the segmentation that are
applied can be found in \autoref{tab:pitches}. The path length of the particle
is then computed for each segment that the particle crosses, and the sensor
thickness is taken into account. A charge deposit is calculated based on the
path length of the particle and particle properties such as its charge and
momentum. The calculated charge is smeared randomly to emulate measurement noise
effects, and a final per-segment charge threshold is applied.

The cluster position is calculated either as the direct geometric mean
(\emph{digital}, strips) or the charge-weighted mean (\emph{analog}, pixels).
During data processing, clusters retain the full list of activated segments, but
the output dataset only contains the cluster position and variances, which is
a one- or two-dimensional quantity that is constrained to the sensor it is
associated with. To translate to a point in three dimensions, we simply take the centre of the segments in the missing dimension(s). A more precise reconstruction of three-dimensional position (i.e. a ``spacepoint'') will be included in a future release. 

\begin{table}[b]
    \centering
\begin{adjustbox}{center}
    \begin{tabular}{|c|c|c|c|c|c|c|}
         subsystem & barrel pitch [$\mu$m] & barrel id & endcap pitch [$\mu$m] & endcap ids & thickness [$\mu$m] & position type \\
         \hline
         pixel & 50$\times$50 & 17 & 50$\times$50 & 16/18 & 125 & analog \\
         short strips & 80$\times$500 & 24 & 80$\times$500 & 23/25 & 200 & digital \\
         long strips & 100 & 29 & 125 & 28/30 & 250 & digital
    \end{tabular}
\end{adjustbox}
    \caption{Sensor design parameters of the different subsystems in ODD}
    \label{tab:pitches}
\end{table}

\begin{figure}
    \centering
    \begin{subfigure}[t]{0.5\textwidth}
        \includegraphics[width=\linewidth]{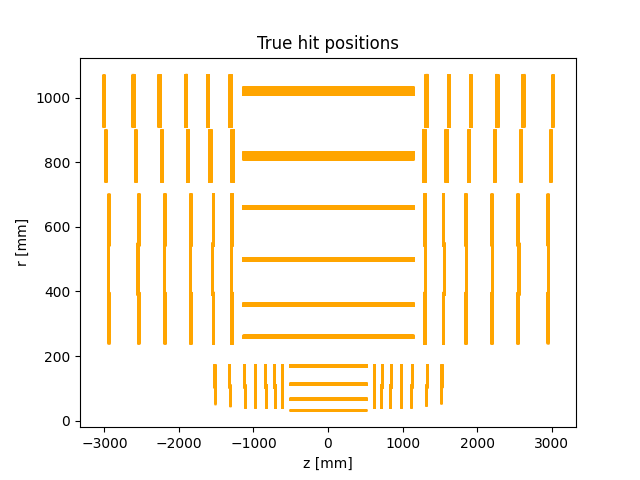}
        \caption{}
        \label{fig:true_hits_pos}
    \end{subfigure}%
    \begin{subfigure}[t]{0.5\textwidth}
        \includegraphics[width=\linewidth]{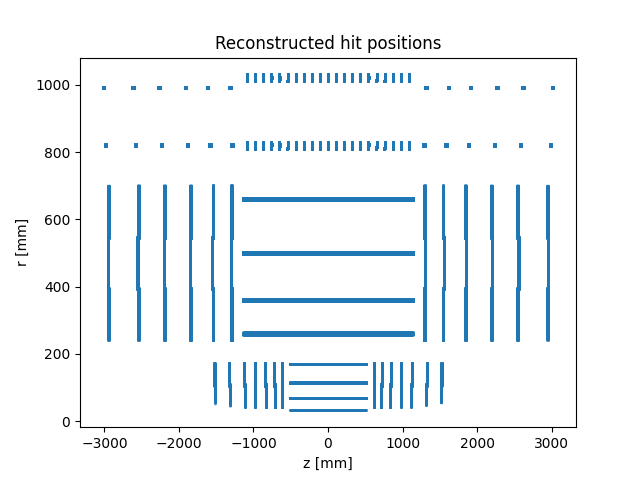}%
        \caption{}
        \label{fig:rec_hits_pos}
    \end{subfigure}
    \caption{Raw hit position (\subref{fig:true_hits_pos}) and reconstructed
    hit position (\subref{fig:rec_hits_pos}) positions in the detector
    demonstrating the effect of pixel, short and long strip digitisation}
    \label{fig:hits_pos}
\end{figure}

\autoref{fig:hits_pos} shows the true and reconstructed positions across the
detector. The true positions are uniformly distributed across all of the sensors,
while the reconstructed positions are discretised to the segmentation according to
the relevant subsystem. Clusters retain the logical connection to the hits that they were created on, which allows navigating back to the original particles from tracks reconstructed
from clusters.

The cluster variances are determined using a statistical approach: for each
sensor category (pixel, short/long strip and barrel, endcap), the residuals
$\Delta l = l_\text{rec} - l_\text{true}$ between the calculated cluster
position and the position of the associated truth hit is computed. This is a
measure of the empirical uncertainty associated with the cluster position
estimate. The RMS of the residual distribution in each measurement dimension
(two for pixel and short-strips, one for long strips) is calculated for each
cluster size. The resulting matrix is stored as a lookup table, and the RMS
values are assigned as variances to the clusters for downstream processing.

\subsection{Calorimeter Digitisation}
\label{sec:cal_digitisation}

The calorimeter cell energy readings in this release are provided with a small set of thresholding effects, to imitate the initial cuts that would be applied in a detector calibration. To begin, a reconstructed time for each contributing particle is approximated by subtracting from the contribution deposit time a time-of-flight correction $dt$ in nanoseconds. This is given as 

\begin{align}
    dt = \sqrt{x^2 + y^2 + z^2} / c - \delta
\end{align}

where $x,y,z$ are the global position of the cell in millimetres, $c=300$\;mm/ns is approximately the speed of light, and $\delta=0.1$ is added as a small buffer to handle time window noise. The particle production times of contributions are required to fall within an absolute $[-1, 10]$\;ns window around the event origin time (taken to be $0$\;ns).

Those contributions that survive are summed per cell, and a very low energy threshold of $50$\;keV ($250$\;keV) is applied to electromagnetic (hadronic) calorimeter cells. These two thresholds reduce the number of particle contributions by around $90\%$ and the number of cells with hits by around $60\%$. 

The calorimeters in the OpenDataDetector are \textit{sampling calorimeters} - by design they register energy deposits relative to their sampling rate. Typically, the energy read-outs would be calibrated by scaling individual cells according to the performance of reconstructed objects. For example, given single-particle simulation, we would expect that the sum of energy deposited across the detector should equal the total energy of the particle. That scaling, while straightforward to extract for EM showers in the OpenDataDetector, requires a proper implementation of particle flow reconstruction for hadronic showers. This will be implemented in future releases of the dataset for both ECal and HCal, alongside the more realistic calorimeter digitisation (for example the inclusion of saturation and noise effects) and reconstruction. As an immediate solution for users interested in calibration, we provide a recipe in \cref{app:adv_calo_digi}, along with resolution of the calibrated calorimeters.

\section{Reconstruction}
\label{sec:reconstruction}

An important component of the \colliderml{} dataset is the ability to use both low-level direct readings from the detector and high-level objects reconstructed using standard algorithms. In this way, practitioners can conveniently use these objects for downstream studies, as well as treating them as a baseline for improved reconstructions. This first release of the dataset includes objects produced as a result of the connection and fitting of inner tracker hits into a particle trajectory (\textit{tracking}). Future releases will include the clustering of calorimeter energy deposits (\textit{clusters} or \textit{topoclustering}), the matching of tracks and clusters characterising single particles (\textit{particle flow}) and the clustering of objects into dense regions believed to originate from a single primary particle (\textit{jet clustering}).

\subsection{Tracking}

Reconstructed tracking objects are produced with the \acts{} Examples framework. \acts{} (A Common Tracking Software) is an experiment independent track reconstruction toolkit, which is used by multiple experiments in production (e.g. ATLAS, Faser, and sPHENIX).

\begin{figure}
    \centering
    \includegraphics[width=0.6\linewidth]{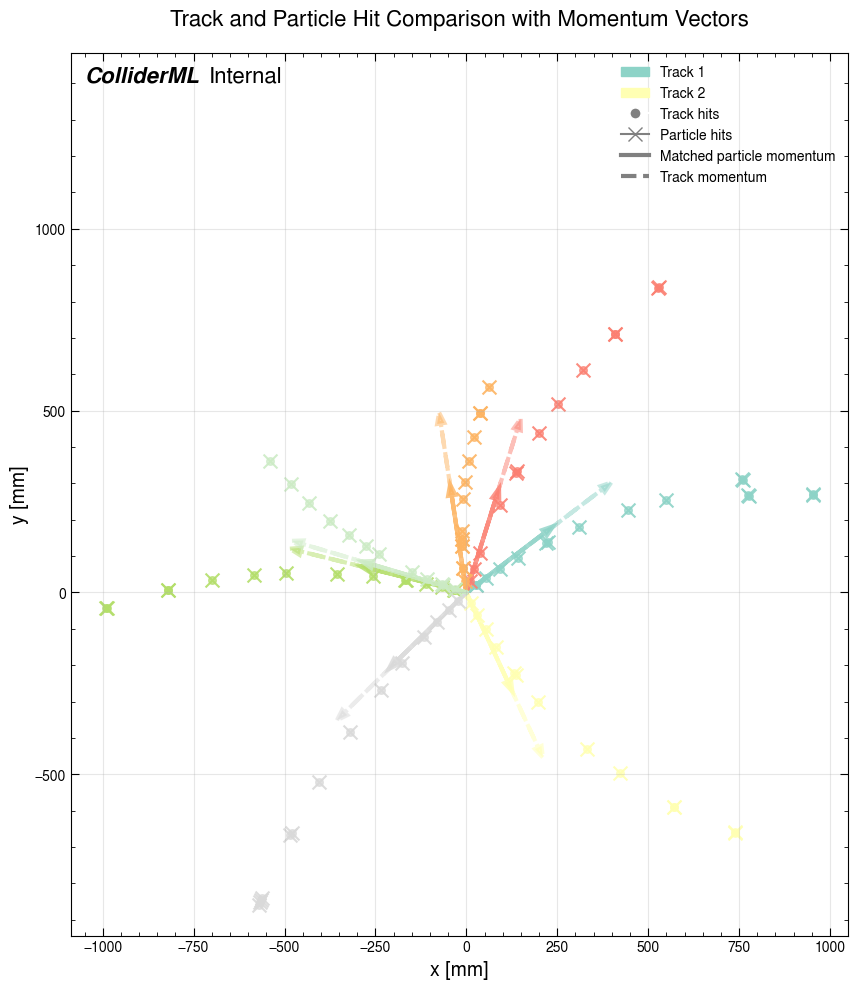}
    \caption{Visualisation of a selection of tracks and their matched particles, and the hits of each. As can be seen, tracks have high completeness (they include most of the hits from their matching particle) and high purity (they are composed mostly of hits from their matching particle). The track parameters are well-fit, such as the initial momentum angle pictured here (the scale of the vector is not meaningful).}
    \label{fig:tracks_and_particles}
\end{figure}

\acts{} provides a set of well known algorithms for different stages of the track reconstruction. Connected-component labelling (CCL) for clustering activations on silicon pixel and strip sensors. The resulting clusters can then be transformed into the global detector frame using space point formation. These space points can be grouped into track seeds and used to get a first estimate of track parameters. A general purpose track finding algorithm based on a combinatorial Kalman filter (CKF) allows to complete potential track seeds to full tracks and to reject fakes by requiring a minimal measurement count. An example set of tracks and particles is visualised in \cref{fig:tracks_and_particles}. 

While the provided \acts{}-based reconstruction chain is not exhaustively tuned, track reconstruction performance is generally satisfying. The \acts{} Examples framework defines various key performance indicators (KPI) which include efficiency, purity, resolutions, and pulls, defined as:

\begin{description}
  \item[Efficiency] Number of reconstructed and uniquely truth matched tracks / number of particles.
  \item[Track content purity] Number of correctly assigned truth hits on track / number of hits on track.
  \item[Resolution] Standard deviation of the Gaussian core of track parameter residuals.
  \item[Pull] Standard deviation of the Gaussian core of track parameter residuals / estimated track parameter uncertainty.
\end{description}

\begin{figure}[htb!]
    \begin{subfigure}[b]{0.8\textwidth}
    \includegraphics[width=1.0\linewidth]{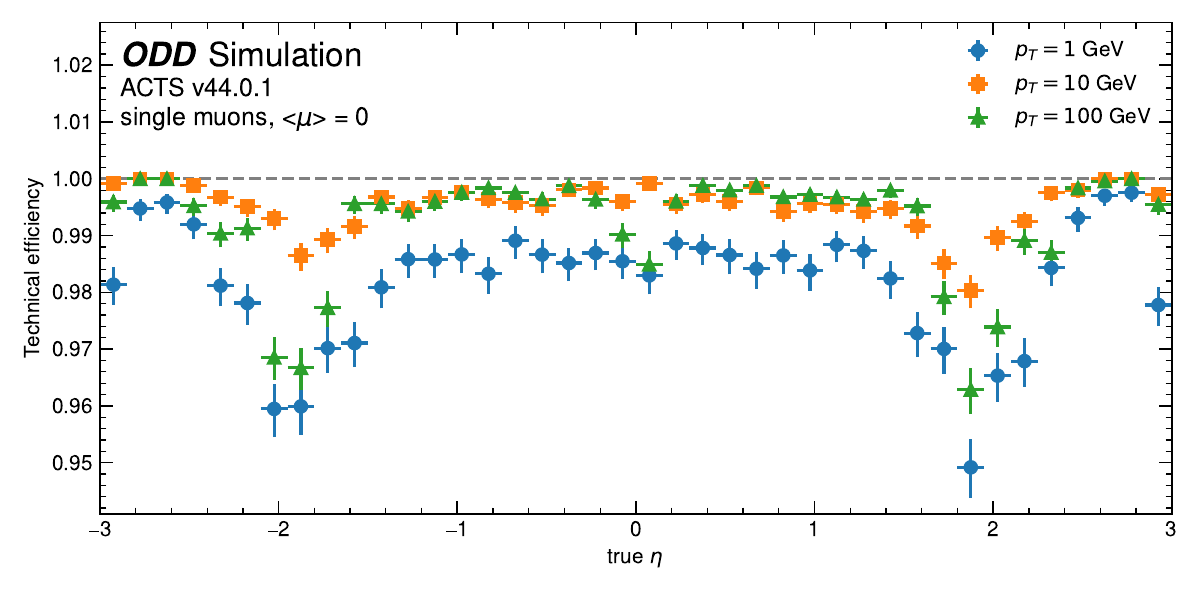}
    \caption{Technical track finding efficiency as a function of true $\eta$ for single $\mu$ with $p_T = 1,10,100$ GeV.}
    \end{subfigure}
    \begin{subfigure}[b]{0.8\textwidth}
    \includegraphics[width=1.0\linewidth]{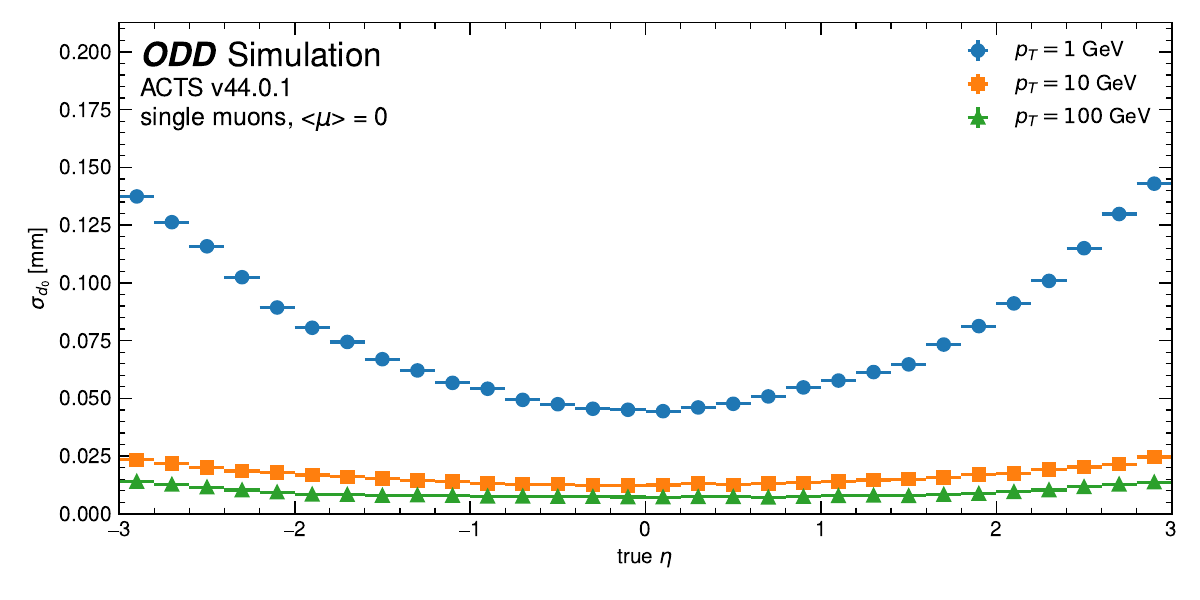}
    \caption{Track resolution of $d_0$ as a function of true $\eta$ for single $\mu$ with $p_T = 1,10,100$ GeV.}
    \end{subfigure}
    \centering
    \begin{subfigure}[b]{0.8\textwidth}
    \includegraphics[width=1.0\linewidth]{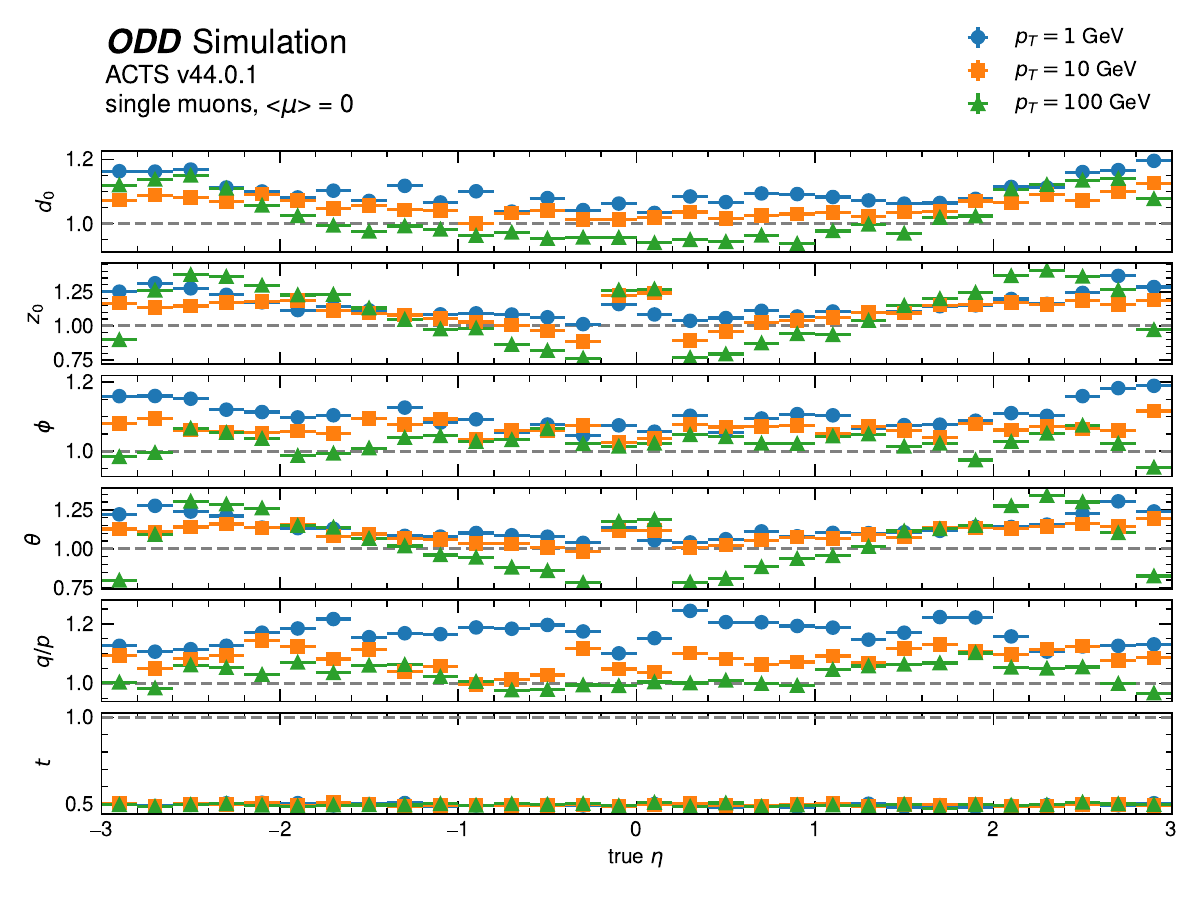}
    \caption{Pull width of all fitted track parameters at the interaction point as a function of true $\eta$ for single $\mu$ with $p_T = 1,10,100$ GeV.}
    \end{subfigure}
    \caption{}\label{fig:muon_tracking_performance}
\end{figure}

\begin{figure}[htb!]
    \centering
    \begin{subfigure}[b]{0.8\textwidth}
    \includegraphics[width=1.0\linewidth]{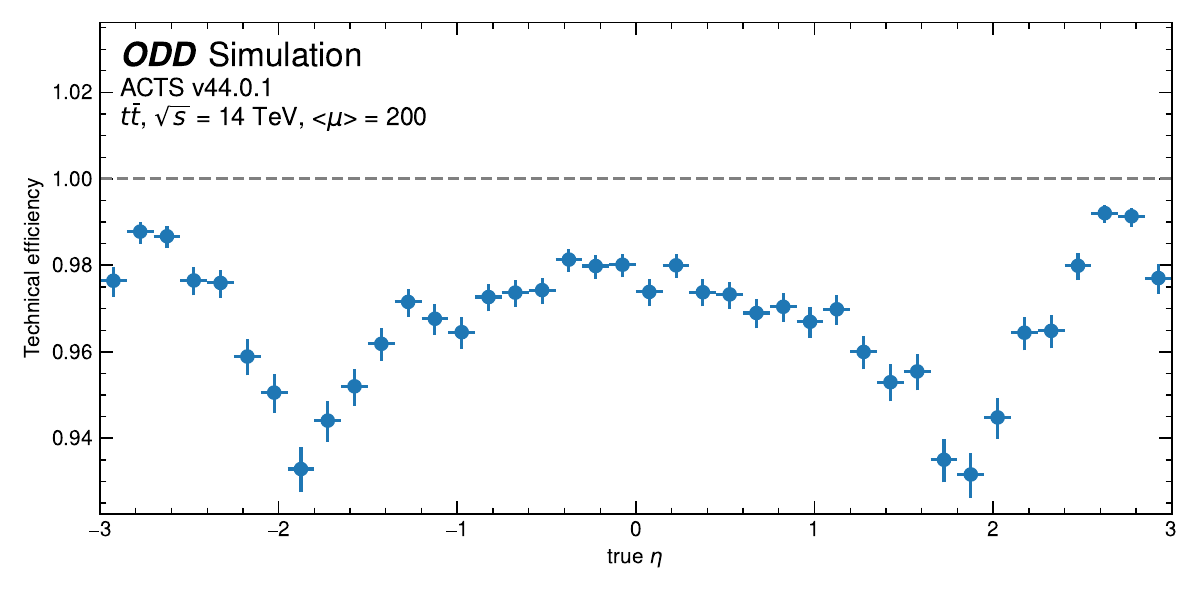}
    \caption{Technical track finding efficiency as a function of true $\eta$ for \ttbar events pileup $\langle \mu \rangle = 200$.}
    \end{subfigure}
    \centering
    \begin{subfigure}[b]{0.8\textwidth}
    \includegraphics[width=1.0\linewidth]{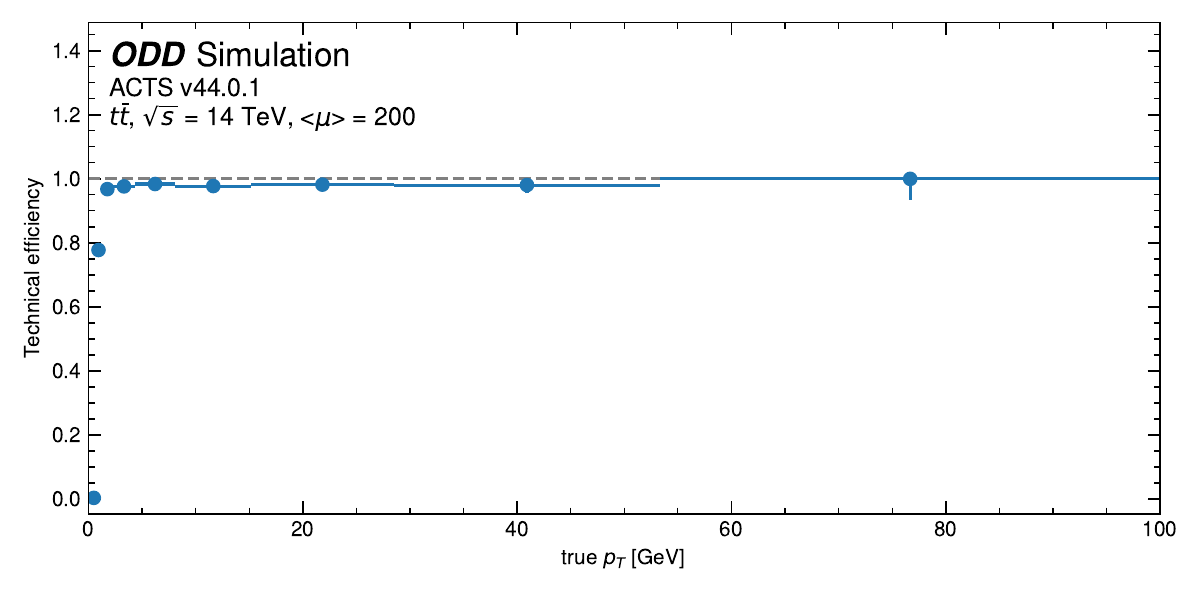}
    \caption{Technical track finding efficiency as a function of true $p_T$ for \ttbar events pileup $\langle \mu \rangle = 200$.}
    \end{subfigure}
    \caption{}\label{fig:ttbar_tracking_performance}
    
\end{figure}

A representative selection of such metrics are reported in \cref{fig:muon_tracking_performance} and \cref{fig:ttbar_tracking_performance}.

\section{Data Structure and Access}
\label{sec:data}

The output of the pipeline described in \cref{fig:release-schedule} is a combination of EDM4hep root files containing simulated inner tracker and calorimeter energy deposits and truth particles, and custom root files containing digitised tracker hits and track candidates. In order to make this collection maximally useful for machine learning experiments, we process this into a set of four parquet files: \texttt{particles} (\cref{tab:particle_structure}), \texttt{tracker\_hits} (\cref{tab:tracker_hits_structure}), \texttt{calo\_hits} (\cref{tab:calo_hits_structure}) and \texttt{tracks} (\cref{tab:tracks_structure}). A quick glance of all object features is given in \cref{tab:data_overview}.

Data is made accessible via the Hugging Face Datasets portal, within the CERN organisation. This location is compatible with the \texttt{datasets} Python library, and therefore a variety of download and streaming options are available through that interface. As an example, to acquire the particle lists of the hard-scatter-only events of the di-Higgs channel, using Python one can call:

\begin{verbatim}
    from datasets import load_dataset
    ds = load_dataset("CERN/ColliderML-Release-1", "dihiggs_pu0_particles")
\end{verbatim}

A convenience library for dataset access and manipulation will be available by running 

\begin{verbatim}
    pip install colliderml
\end{verbatim}

with documentation available at \url{https://opendatadetector.github.io/ColliderML/}. This library contains functions to visualise data, assign daughters to initial state particles, define detector production regions, and downsample events to specific pile-up working points. This last is important for benchmarking purposes. We propose that results obtained on \colliderml{} should, by default, be tested against a pile-up of 200 interactions per bunch crossing. However, it is likely to be useful for development studies to quote performance on hard-scatter-only, Run 1 conditions (pile-up 20), Run 2 conditions (pile-up 40) or Run 3 conditions (pile-up 60). This allows for scaling behaviour to be examined across increasing luminosity. Future work will explore rigorous benchmarking metrics and techniques on this dataset.

\begin{table}[H] 
    \centering
    \caption{Overview of physics objects in Release 1 and their features}
    \label{tab:data_overview}
    \begin{tabularx}{\textwidth}{@{}l X@{}} 
        \toprule
        \textbf{Object} & \textbf{Key Features} \\ 
        \midrule
        \textbf{Particles} & 
        \textbf{Kinematics:} 4-momentum ($p_x, p_y, p_z, E$), mass, charge. \newline
        \textbf{Vertex:} Production pos ($\vec{v}, t$) and Truth Perigee ($d_0, z_0$). \newline
        \textbf{Identity:} PDG code, Unique ID, Parent ID. \newline
        \textbf{Origin:} Primary particle boolean, Primary vertex index. \\ 
        \midrule
        \textbf{Tracker Hits} & 
        \textbf{Position:} Measured ($x,y,z$) vs. True ($x,y,z$), Time. \newline
        \textbf{Hierarchy:} Volume, Layer, Surface, and Detector IDs. \newline
        \textbf{Truth:} Link to generating \texttt{particle\_id}. \\ 
        \midrule
        \textbf{Calorimeter Hits} & 
        \textbf{Energy:} Total energy and Cell position ($x,y,z$). \newline
        \textbf{Contributions:} Nested lists of truth particles contributing to the cell (IDs, energy fractions, timing). \\ 
        \midrule
        \textbf{Tracks} & 
        \textbf{Fitted Helix:} $d_0, z_0, \phi, \theta, q/p$. \newline
        \textbf{Associations:} List of tracker \texttt{hit\_ids}. \newline
        \textbf{Truth:} \texttt{majority\_particle\_id} (truth matching). \\ 
        \bottomrule
    \end{tabularx}
\end{table}

\section{Summary and Outlook} \label{sec:conclusion}

In this note, we have introduced \colliderml{}, a large-scale, open, and experiment-agnostic dataset designed to bridge the gap between fast-simulation benchmarks and the realistic, high-pileup conditions of the HL-LHC. By providing one million fully simulated events with realistic digitisation across ten physics channels , \colliderml{} offers a unique testbed for developing and stress-testing the next generation of classical and machine learning algorithms.

This first release focuses on the lowest-level objects—energy deposits, tracks and truth particles—enabling immediate work on tracking and calorimetry calibration. Future releases are planned to expand this ecosystem significantly. Release 2 will introduce higher-level reconstruction objects, including particle flow candidates and jets, to facilitate cross-stage studies. Future releases will introduce realistic detector miscalibrations and ``challenge'' datasets, specifically designed to probe the robustness and domain-adaptation capabilities of modern ML architectures.

We make this data available on Hugging Face to lower the barrier to entry for the ML community. We invite practitioners to use \colliderml{} not just as a static benchmark, but as a dynamic environment to study the fundamental limits of reconstruction and generation in high-luminosity collider environments, and we invite contributions to metric definitions and reconstruction approaches.

\section{Acknowledgements}
\label{sec:ack}

DM is supported by the Danish Data Science Academy, which is funded by the Novo Nordisk Foundation (NNF21SA0069429). PG is supported by the Eric \& Wendy Schmidt Fund for Strategic Innovation through the CERN Next Generation Triggers project under grant agreement number SIF-2023-004. This research used resources of the National Energy Research Scientific Computing Center, a DOE Office of Science User Facility supported by the Office of Science of the U.S. Department of Energy under Contract No. DE-AC02-05CH11231 using NERSC award HEP-ERCAP0034031.

\bibliography{colliderml,datasets,ml_models}

\newpage

%%%%%%%%%%%%%%%%%%%%%%%%%%%%%%%%%%%%%%%%%%%%%%%%%%%%%%%%%%%%

\appendix

\section{Appendix / supplemental material}

\subsection{DD4hep Multithreading Scaling Behaviour}
\label{app:dd4hep_mt}

\begin{figure}[!h]
    \centering
    \includegraphics[width=0.8\linewidth]{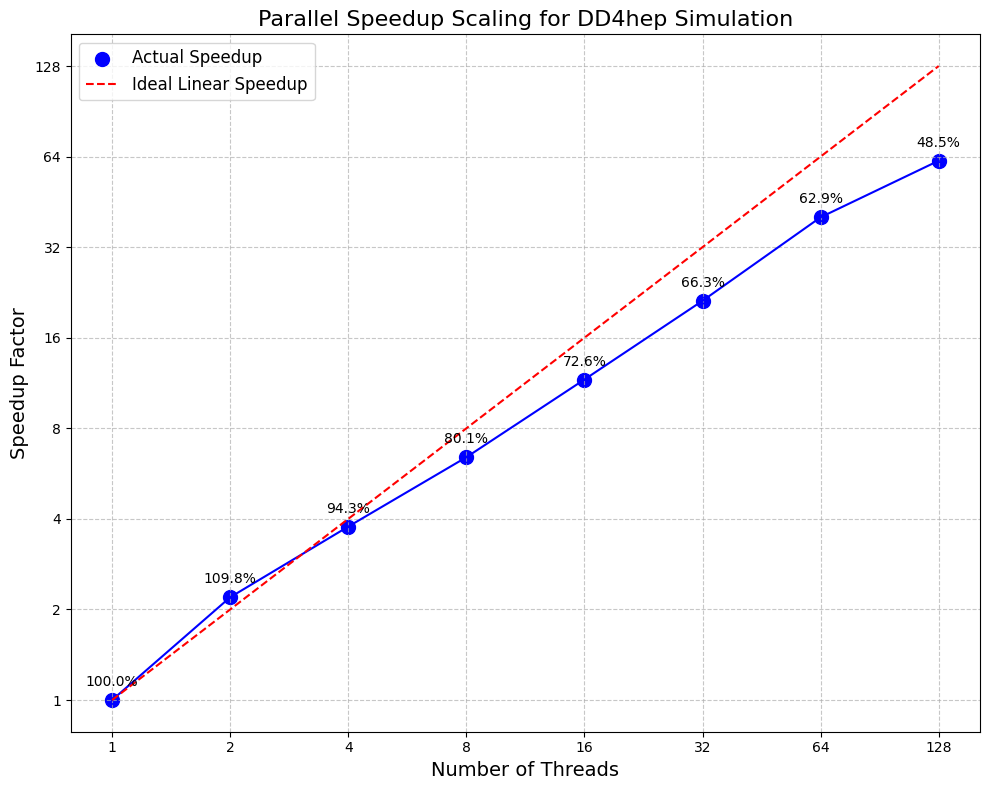}
    \caption{Scaling performance of multithreaded DD4hep. Each thread simulates 8 events of $t\bar{t}$ at $\langle \mu \rangle = 10$. Speed-up factor is given as the number of threads times the ratio of single-threaded wall-time to multi-threaded wall-time at that thread-count. Included are the percentages of measured speed-up compared with ideal strong-scaling speed-up.}
    \label{fig:dd4hep_mt}
\end{figure}

A technically significant development of this dataset is that it is the first to use large-scale multithreading in the DD4hep simulation framework. This proved essential to being able to provide the number of events currently available. To understand why, note that the previous solution to parallelisation required invoking an independent \texttt{ddsim} instance on each core, which rapidly consumed the available memory of a compute node. As such, memory limitations capped multi-processing of simulations to 32 per compute node. With multi-threading, we see good scaling up to 8 parallel threads, with the possibility of 128 threads per node, observed in \cref{fig:dd4hep_mt}.

\subsection{DD4hep Simulation Truth Handling}
\label{app:truth_handling}

To enable sophisticated studies of particle decay reconstruction or simulation, we enforce very conservative truth handling. The DD4hep algorithm proceeds as: In a particle decay/interaction, should we retain the daughter information? If yes, then the daughter is registered as a particle with all its details entering the particle record and all its energy deposits attributed to it. If no, then the daughter is marked to be discarded, and its energy deposits are attributed to its parent. This attribution occurs after the simulation, and thus we may have the case where a to-be-discarded daughter itself produces a to-be-retained grand-daughter. As expected then, the daughter's hits are attributed to its parent, while the grand-daughter's hits are attributed to itself, and so on. Given this framework, the choice of DD4hep truth handling amounts to: What are the conditions in which a particle is discarded or retained?

\begin{figure}[!htbp]
\begin{subfigure}[b]{0.42\textwidth}
    \centering
    \includegraphics[width=0.8\linewidth]{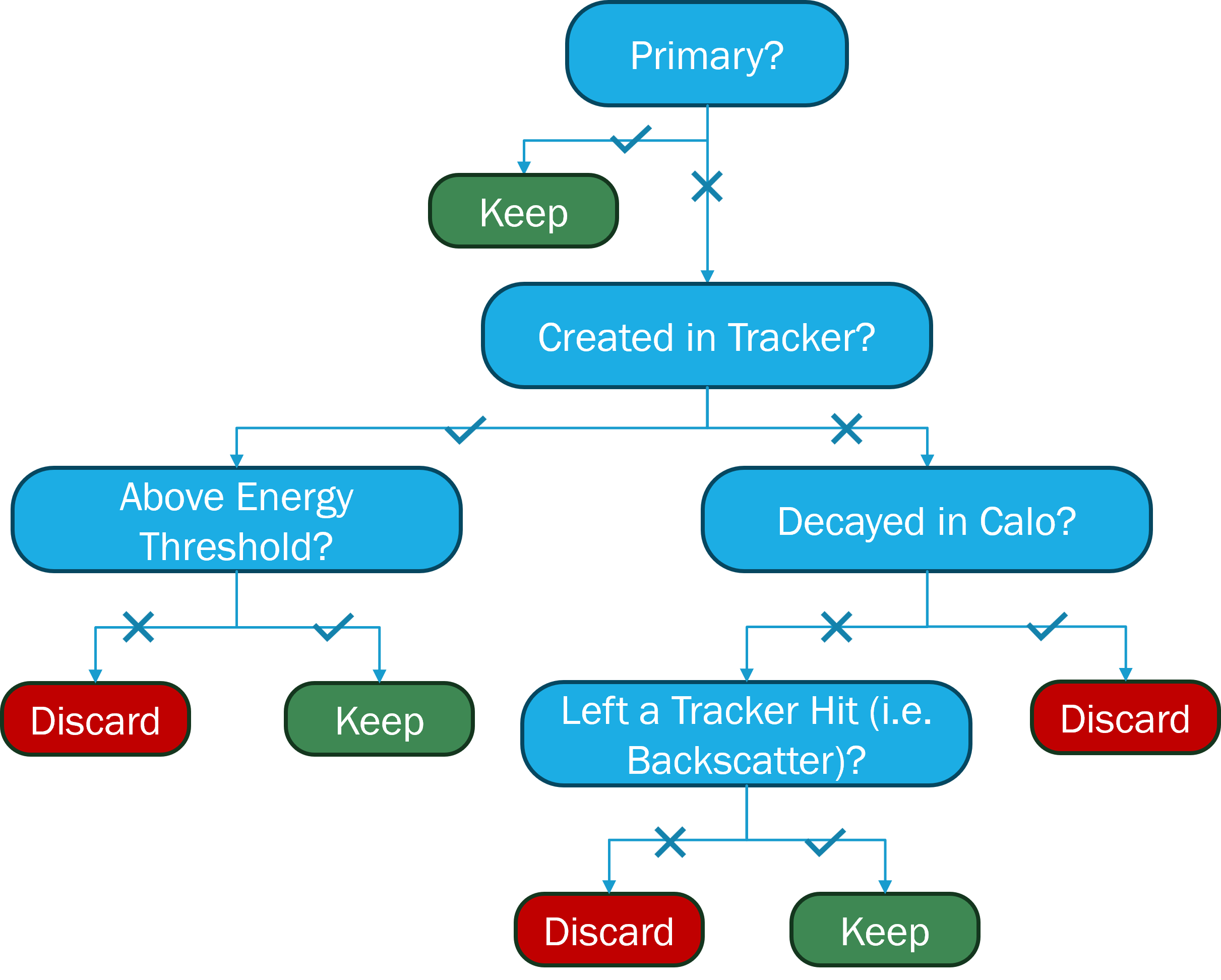}
    \caption{}
        \label{fig:default_truth_handling}
\end{subfigure}
\begin{subfigure}[b]{0.47\textwidth}
        \centering
    \includegraphics[width=0.95\linewidth]{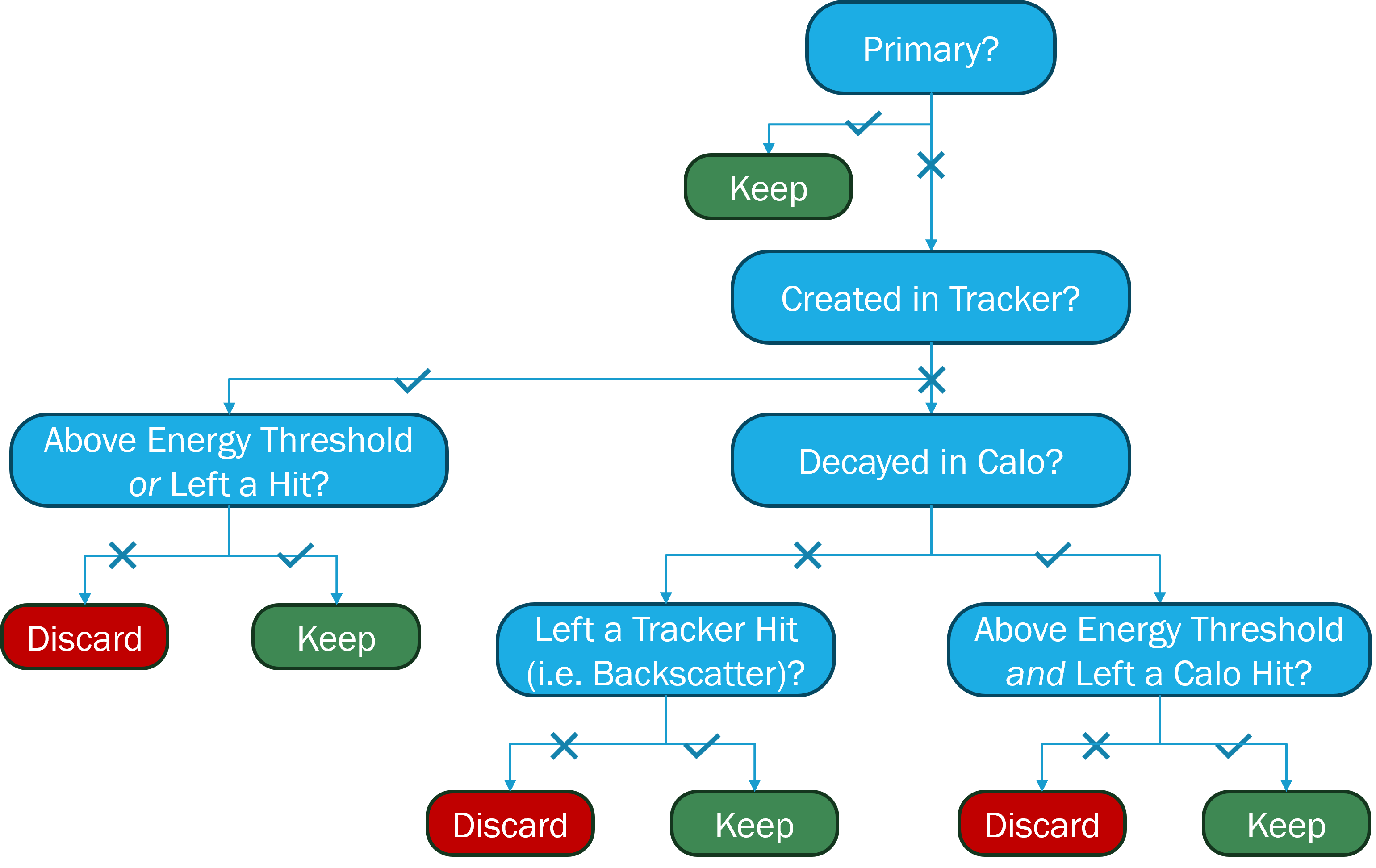}
    \caption{}
    \label{fig:dd4hep_full_truth_handling}
    \end{subfigure}
    \caption{Truth handling logical flow. a) The default DD4hep truth handling. b) The custom \colliderml handling, allowing more detailed tracking of particle information through the detector}
    
\end{figure}

The default truth handling is shown in \cref{fig:default_truth_handling}. While this leads to compact particle truth records, and is likely suitable for the majority of standard analysis use-cases, it leads to a variety of behaviour that is non-ideal. For example, backscatter from calorimeter into the tracker is not correctly handled. Additionally, tracking can become confusing in this environment since all low energy particles are assigned to their parent, and thus their hits will appear to create tree-like tracks in the tracker. Finally, all calorimeter decay information is lost.

\subsection{Calorimeter Calibration}
\label{app:adv_calo_digi}

We present a simple recipe here for calibration of the OpenDataDetector calorimeters. The energy deposited in each cell in the \colliderml{} release is already thresholded, so one needs to apply the correction factors described in \cref{tab:calo_calibration} according to the location of the cell.

\begin{table}[]
    \centering
    \renewcommand{\arraystretch}{1.7} % Adjust row spacing
    \begin{tabular}{l|l}
    \hline
         \textbf{Region} & \textbf{Scaling factor}\\ \hline
         ECal Barrel & $37.5$\\
         ECal Endcap & $38.7$\\
         HCal Barrel & $45.0$\\
         HCal Endcap & $46.9$\\
         \hline
    \end{tabular}\vspace{5pt}
    \caption{Current estimate of the scaling factors for the Open Data Detectors. The numbers may change (especially for the hadronic calorimeter) once the full reconstruction procedure is implemented and showers are fully calibrated.}
    \label{tab:calo_calibration}
\end{table}

\begin{figure}[!htbp]
    \centering
    % First row
    \begin{subfigure}[b]{0.48\textwidth}
        \centering
        \includegraphics[width=\linewidth]{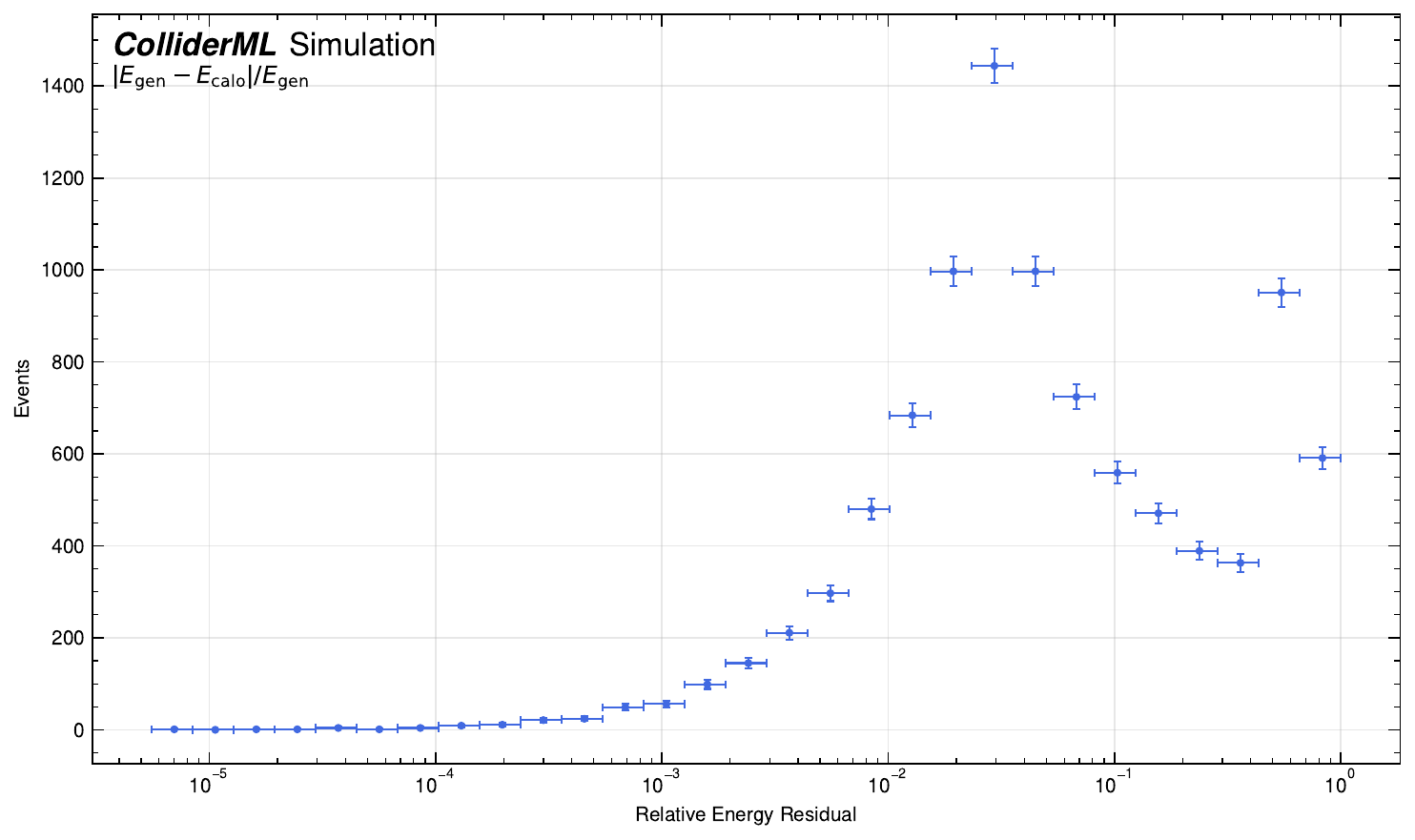}
        \caption{}
        \label{fig:electron_residual_relative}
    \end{subfigure}
    \hfill
    \begin{subfigure}[b]{0.48\textwidth}
        \centering
        \includegraphics[width=\linewidth]{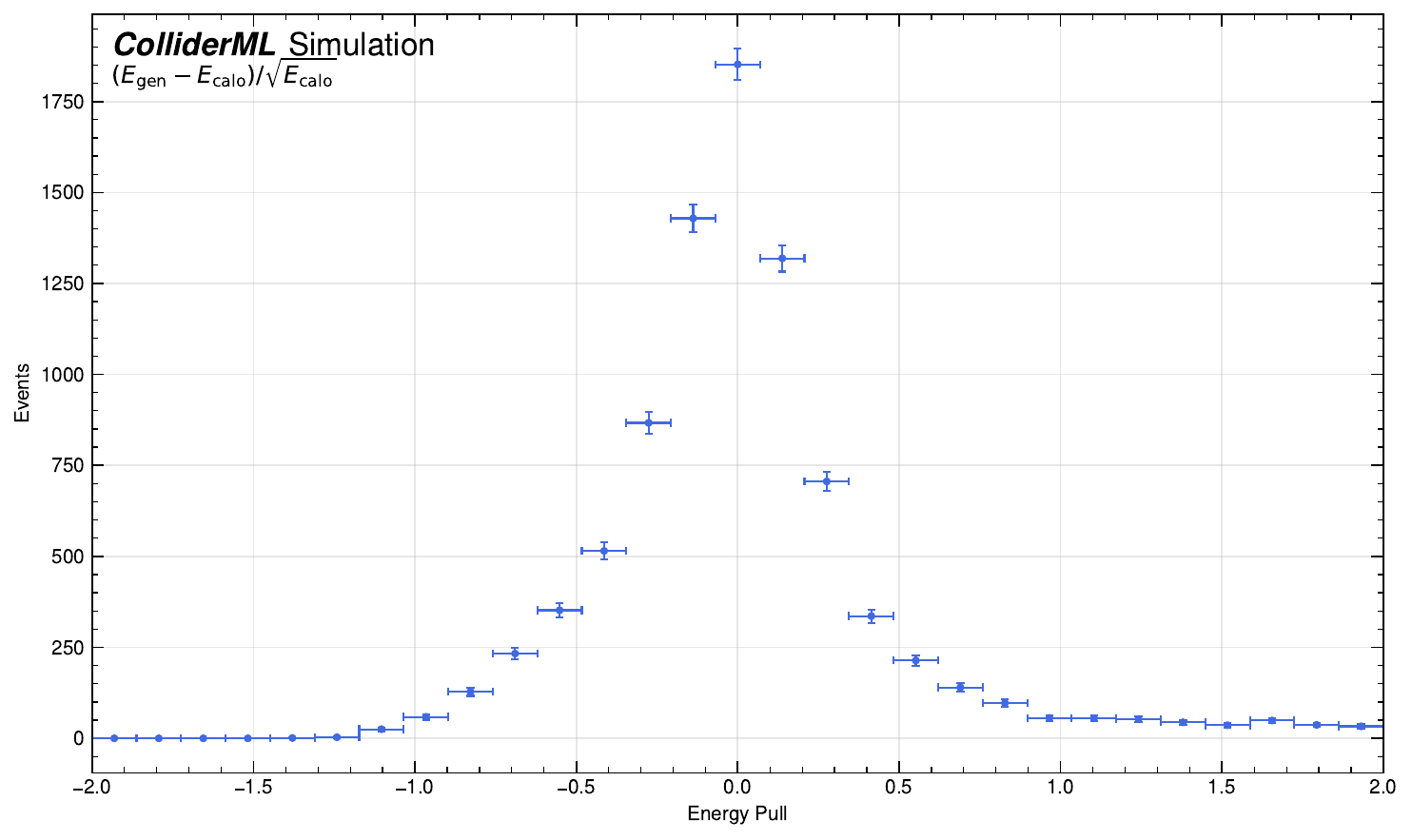}
        \caption{}
        \label{fig:electron_pull}
    \end{subfigure}
    
    \vspace{1em} % small vertical space between rows
    
    % Second row
    \begin{subfigure}[b]{0.48\textwidth}
        \centering
        \includegraphics[width=\linewidth]{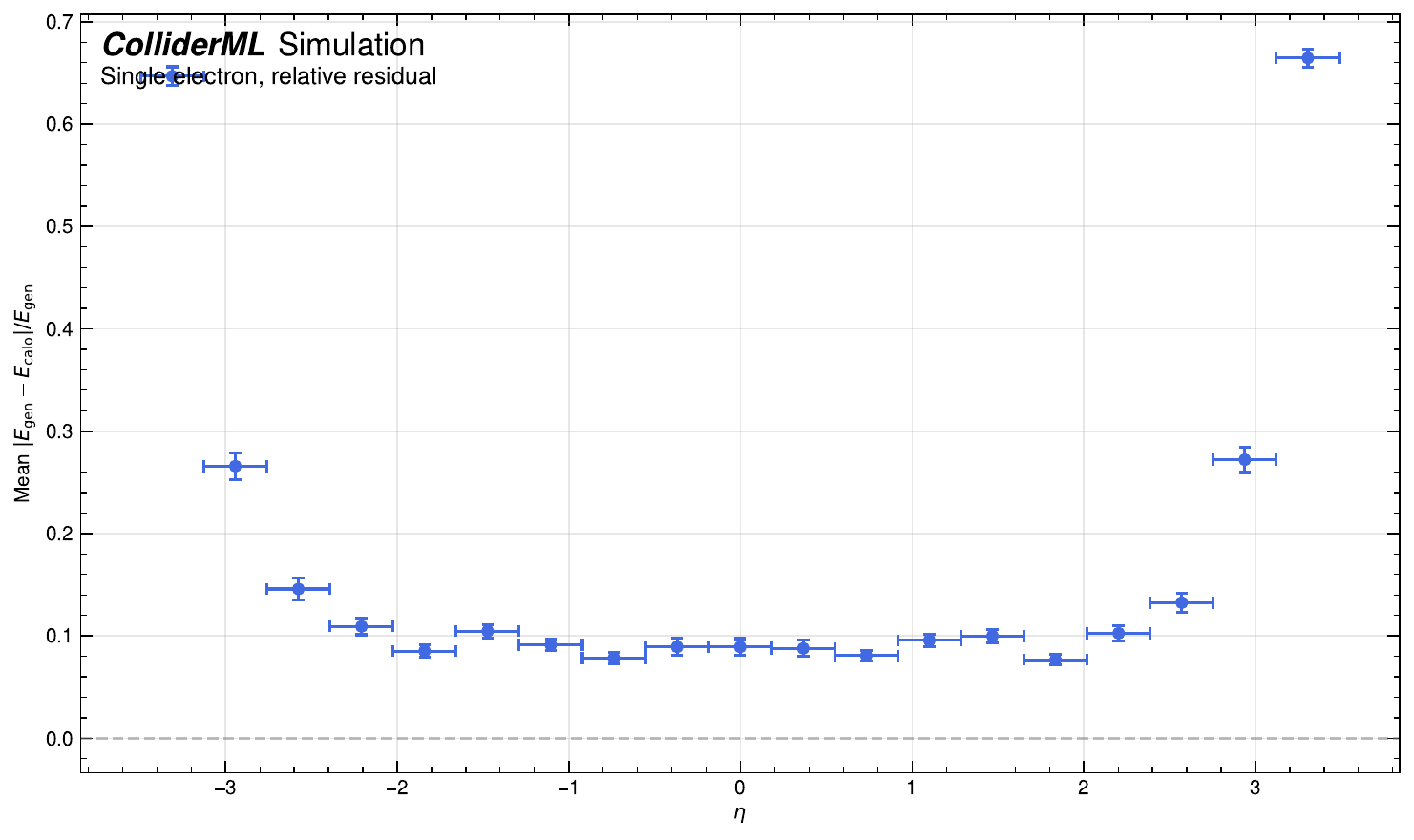}
        \caption{}
        \label{fig:electron_profile_eta}
    \end{subfigure}
    \hfill
    \begin{subfigure}[b]{0.48\textwidth}
        \centering
        \includegraphics[width=\linewidth]{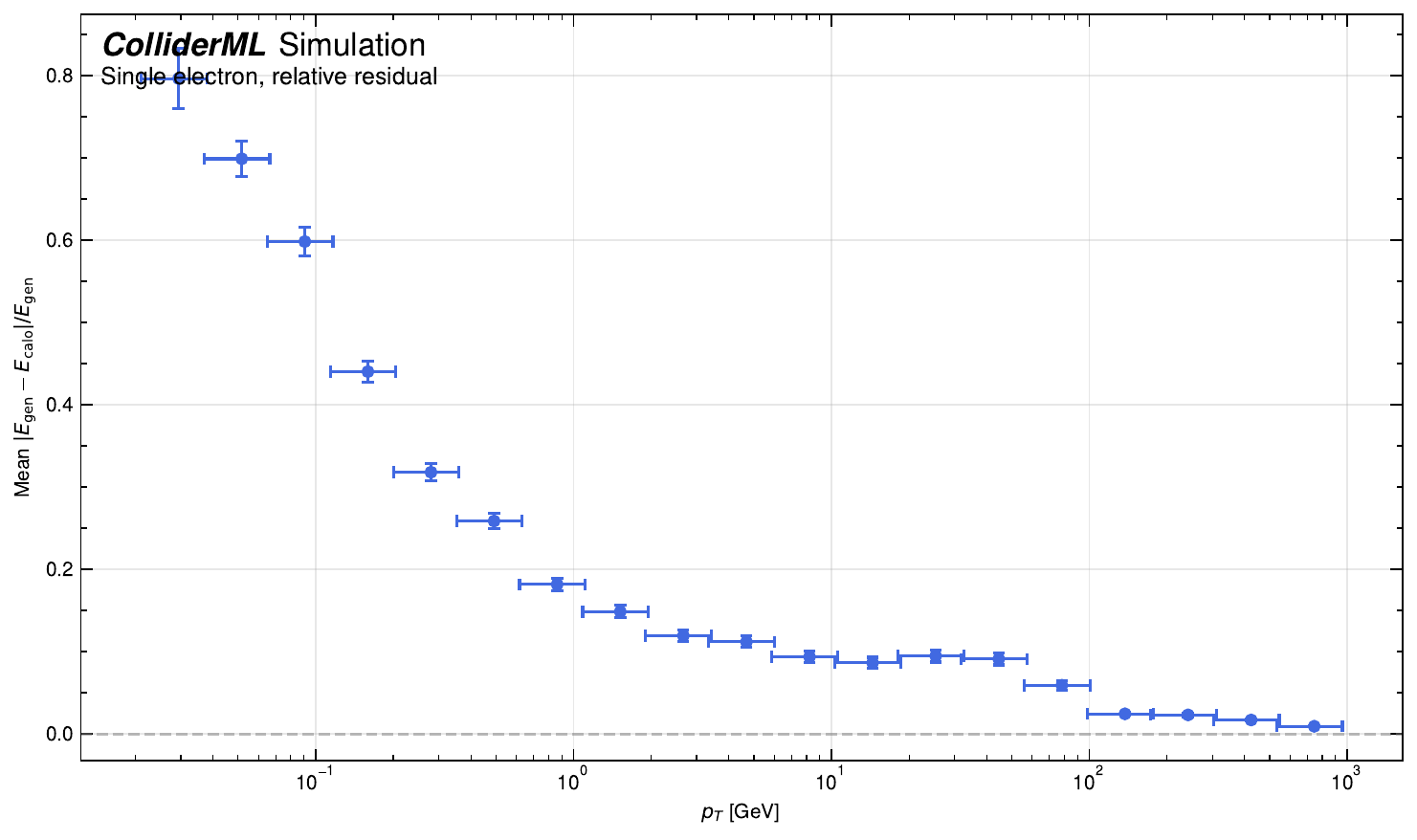}
        \caption{}
        \label{fig:electron_profile_pt}
    \end{subfigure}

    \caption{Calorimeter digitisation validation results for single electron samples: a) Relative residual distribution, b) Pull distribution, c) Relative residual across pseudorapidity, and d) Relative residual across transverse momentum of the generator particle}
    \label{fig:calo_resolution_electron}
\end{figure}

\begin{figure}[!htbp]
    \centering
    % First row
    \begin{subfigure}[b]{0.48\textwidth}
        \centering
        \includegraphics[width=\linewidth]{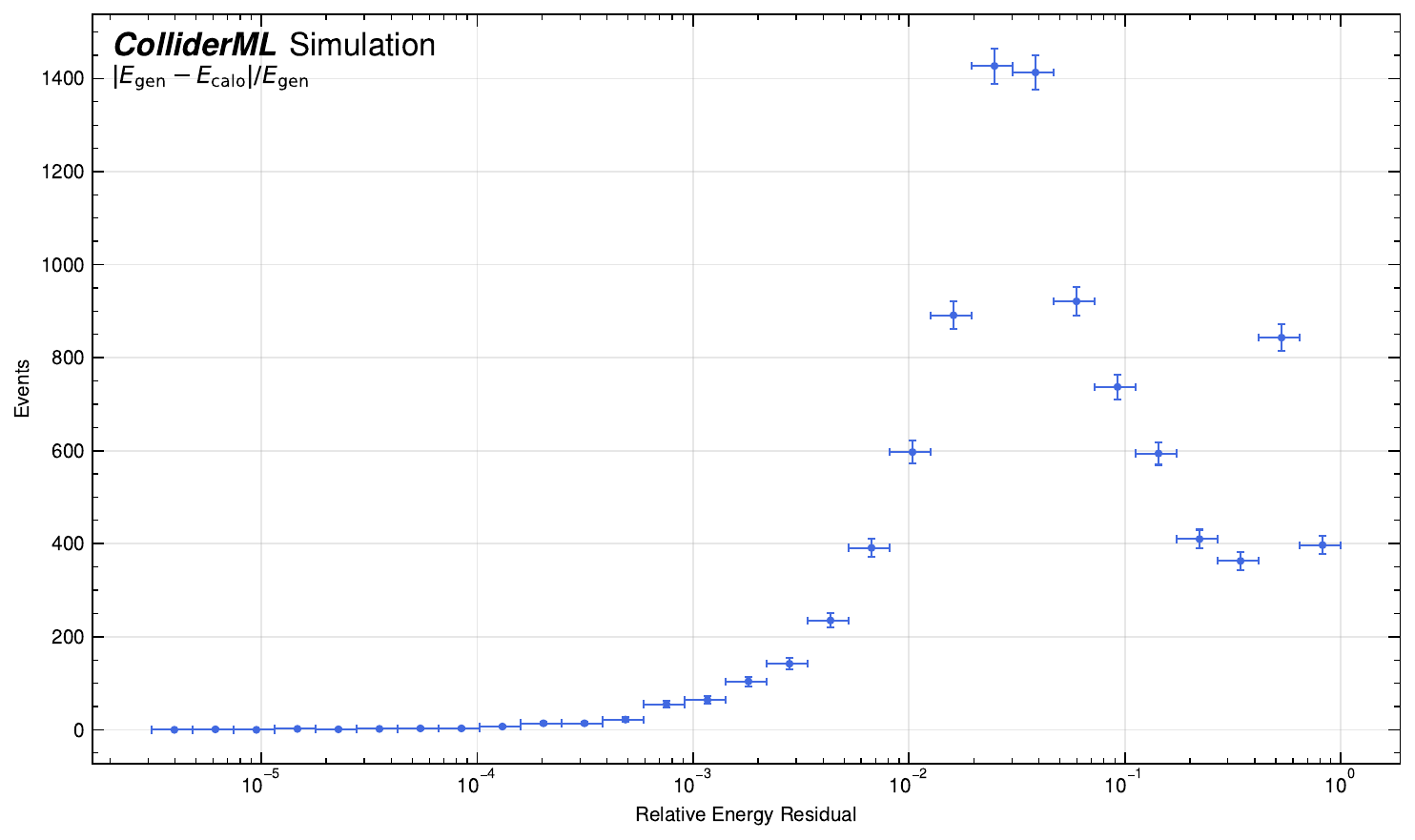}
        \caption{}
        \label{fig:photon_residual_relative}
    \end{subfigure}
    \hfill
    \begin{subfigure}[b]{0.48\textwidth}
        \centering
        \includegraphics[width=\linewidth]{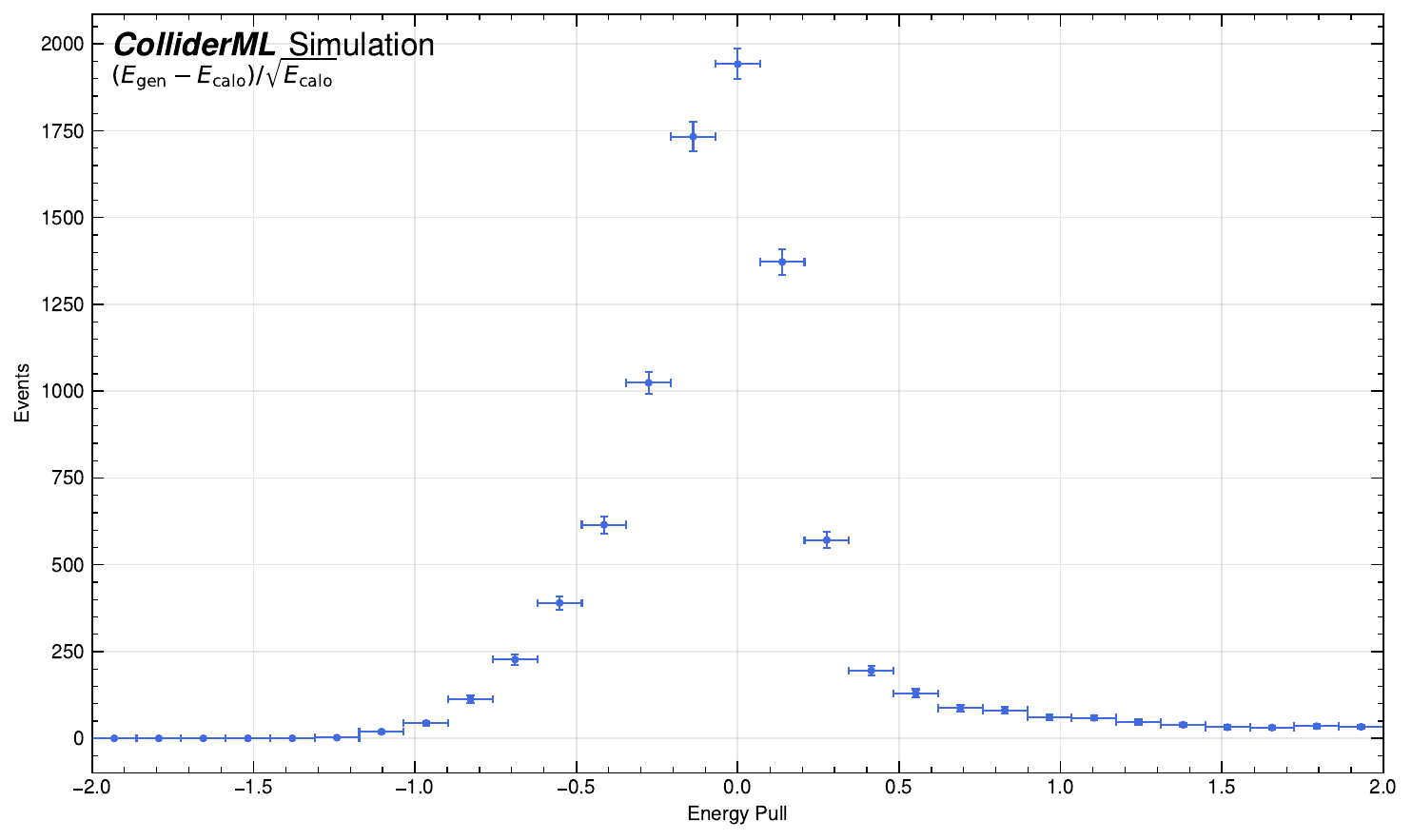}
        \caption{}
        \label{fig:photon_pull}
    \end{subfigure}
    
    \vspace{1em} % small vertical space between rows
    
    % Second row
    \begin{subfigure}[b]{0.48\textwidth}
        \centering
        \includegraphics[width=\linewidth]{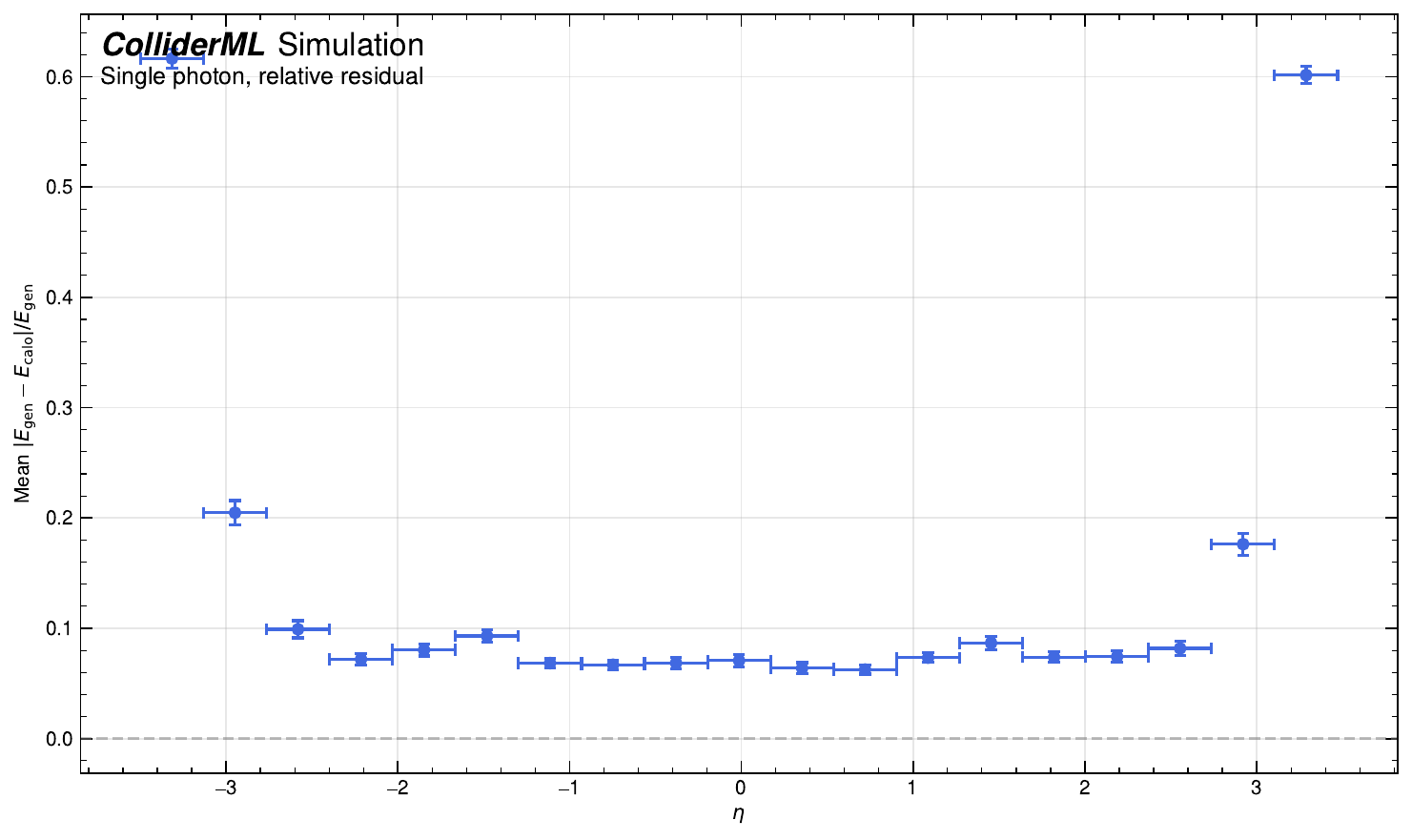}
        \caption{}
        \label{fig:photon_profile_eta}
    \end{subfigure}
    \hfill
    \begin{subfigure}[b]{0.48\textwidth}
        \centering
        \includegraphics[width=\linewidth]{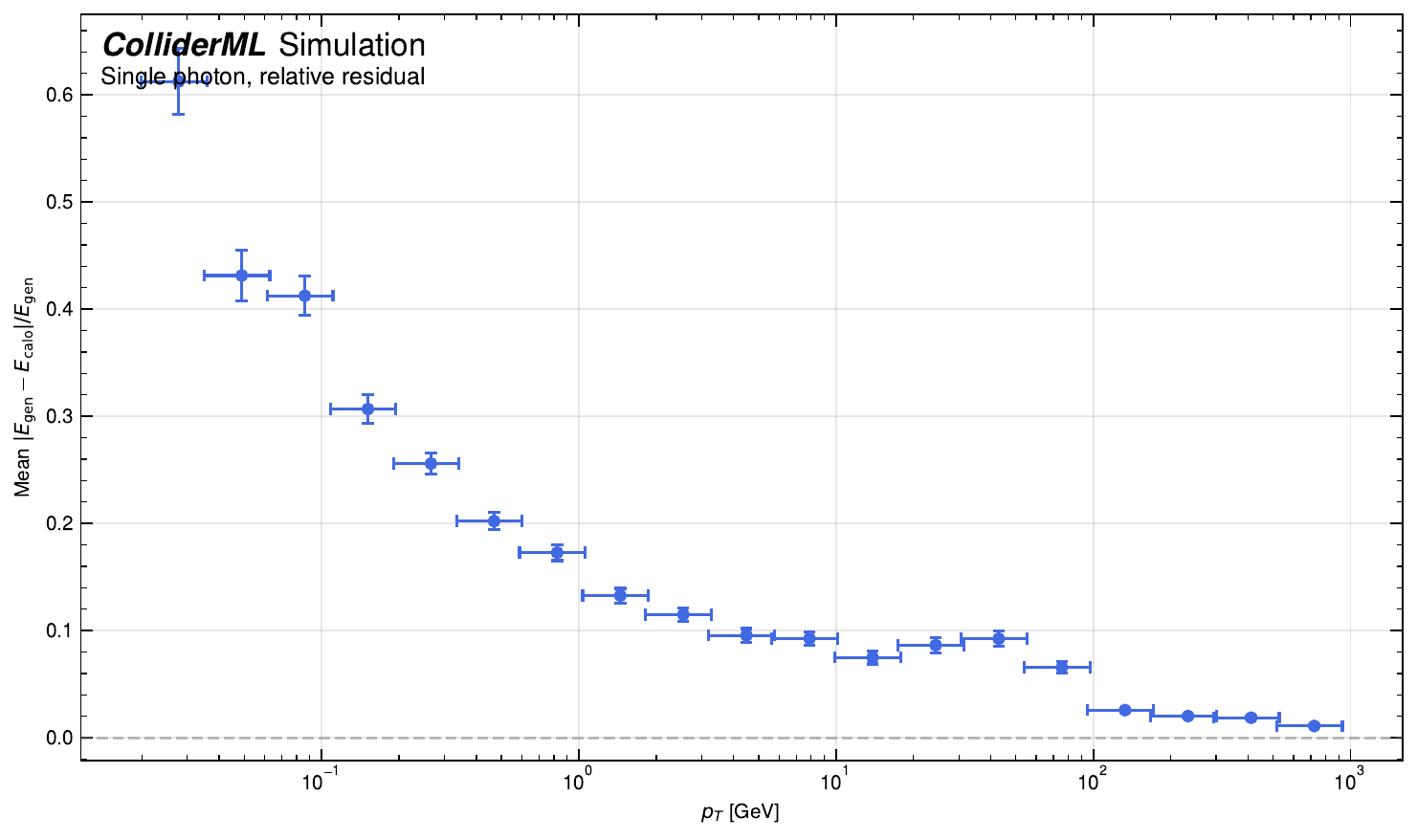}
        \caption{}
        \label{fig:photon_profile_pt}
    \end{subfigure}

    \caption{Calorimeter digitisation validation results for single photon samples: a) Relative residual distribution, b) Pull distribution, c) Relative residual across pseudorapidity, and d) Relative residual across transverse momentum of the generator particle}
    \label{fig:calo_resolution_photon}
\end{figure}

\begin{figure}[!htbp]
    \centering
    % First row
    \begin{subfigure}[b]{0.48\textwidth}
        \centering
        \includegraphics[width=\linewidth]{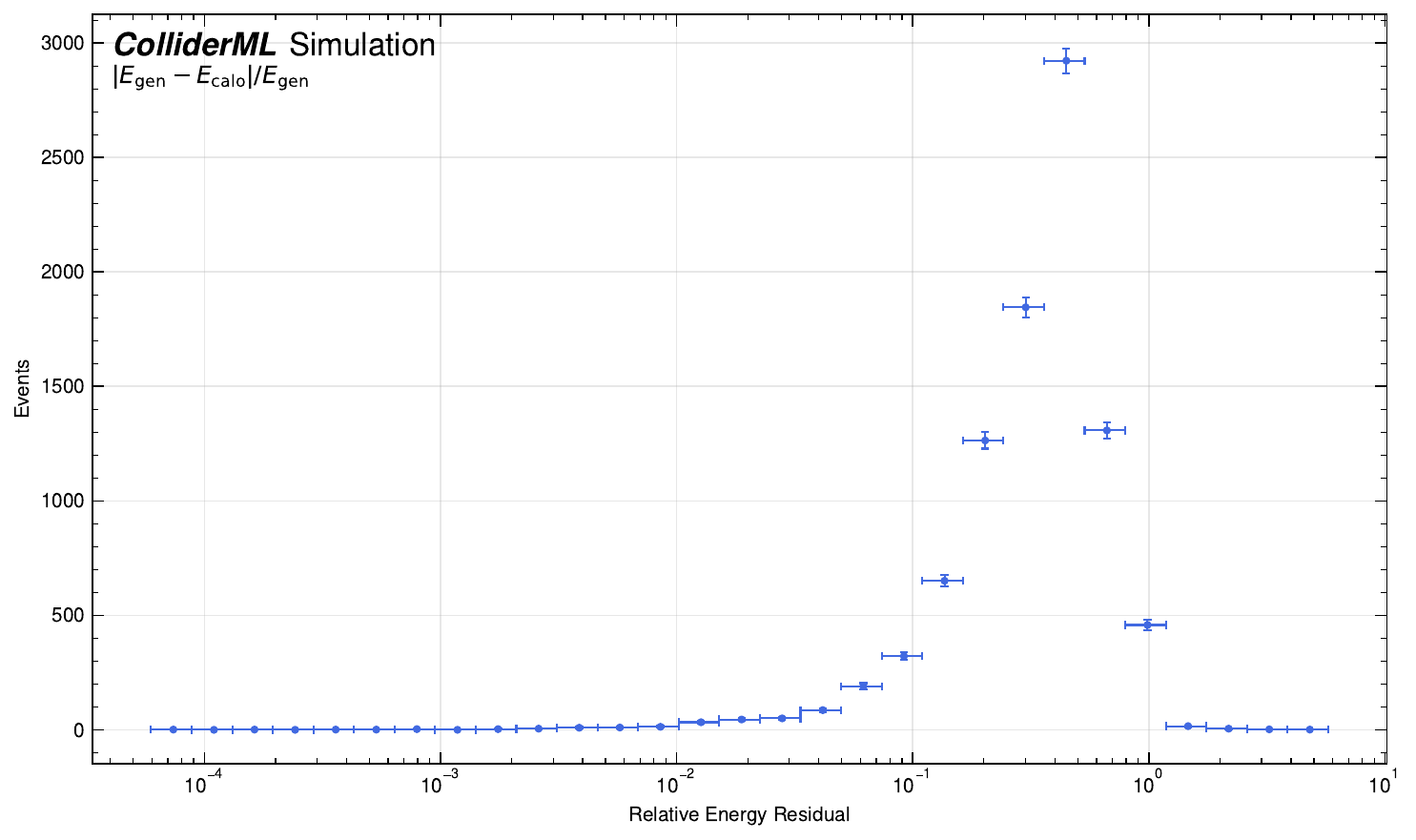}
        \caption{}
        \label{fig:pion_residual_relative}
    \end{subfigure}
    \hfill
    \begin{subfigure}[b]{0.48\textwidth}
        \centering
        \includegraphics[width=\linewidth]{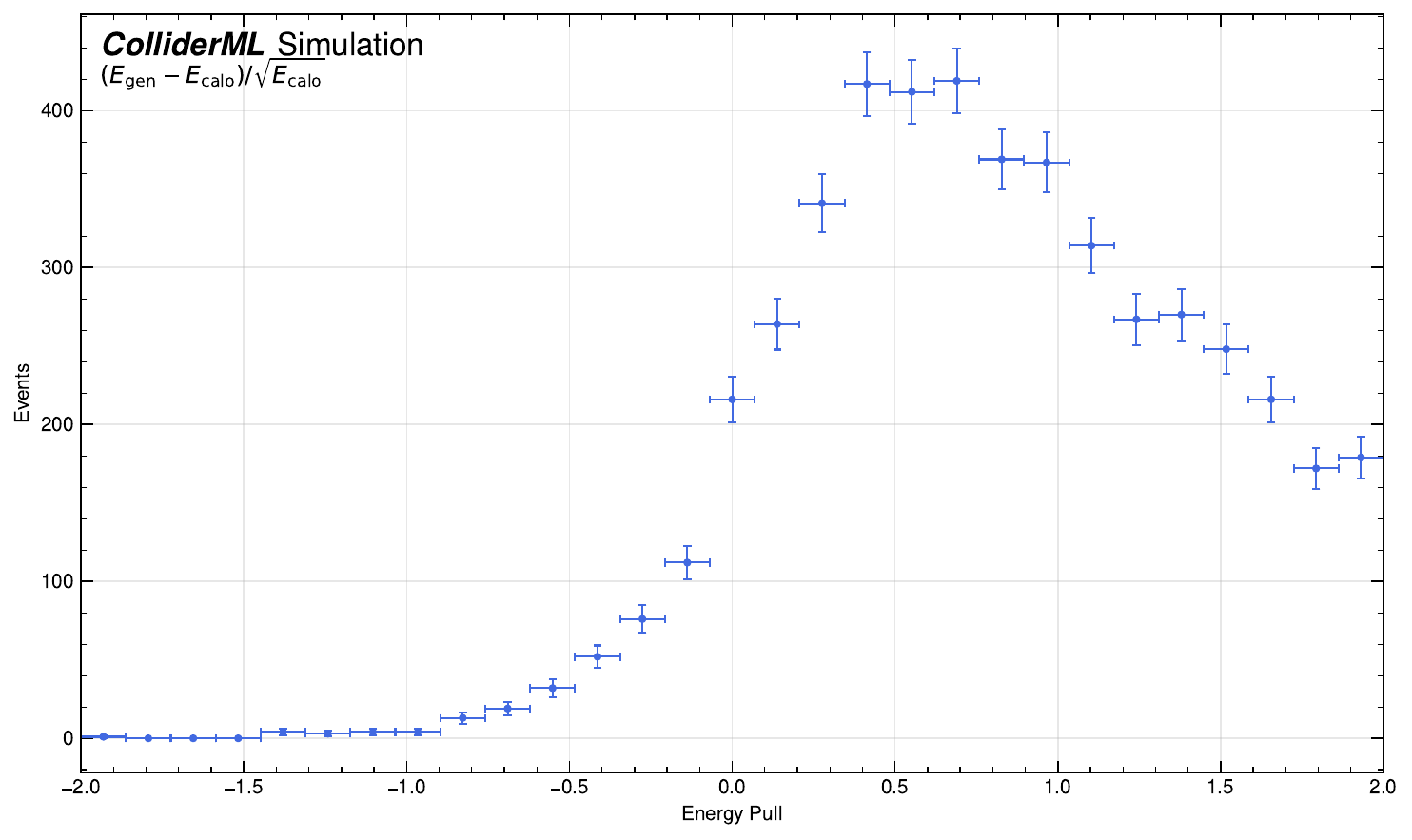}
        \caption{}
        \label{fig:pion_pull}
    \end{subfigure}
    
    \vspace{1em} % small vertical space between rows
    
    % Second row
    \begin{subfigure}[b]{0.48\textwidth}
        \centering
        \includegraphics[width=\linewidth]{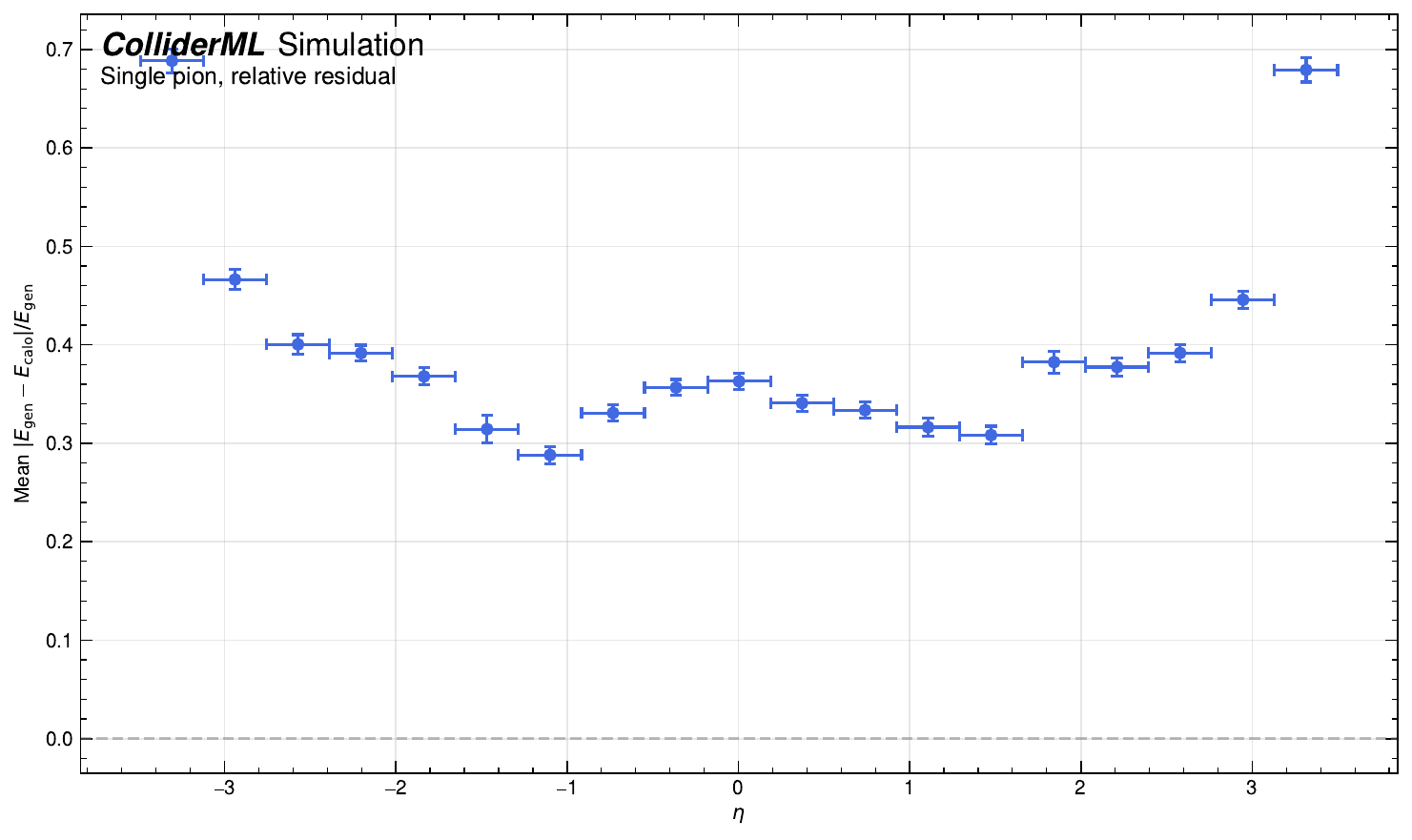}
        \caption{}
        \label{fig:pion_profile_eta}
    \end{subfigure}
    \hfill
    \begin{subfigure}[b]{0.48\textwidth}
        \centering
        \includegraphics[width=\linewidth]{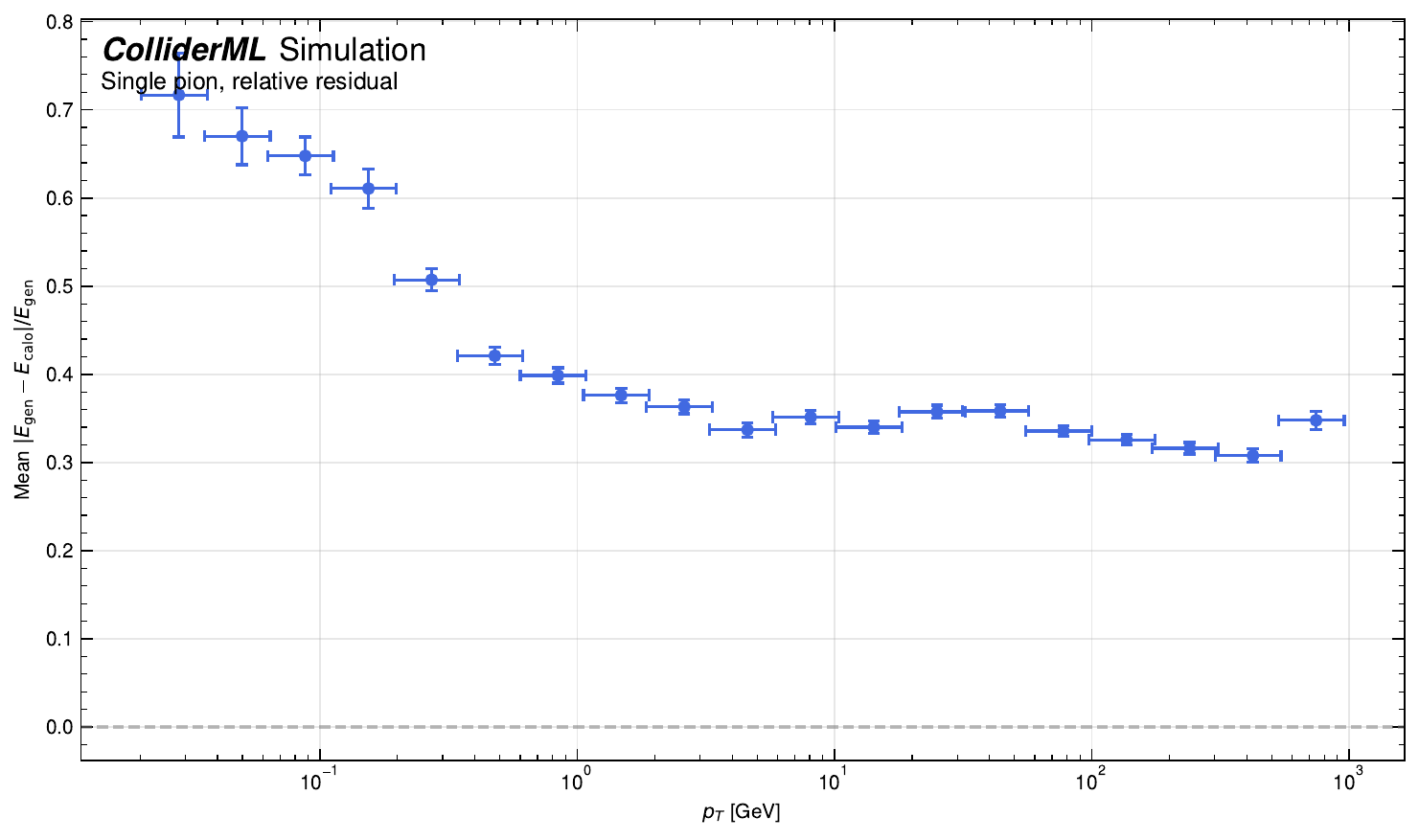}
        \caption{}
        \label{fig:pion_profile_pt}
    \end{subfigure}

    \caption{Calorimeter digitisation validation results for single pion samples: a) Relative residual distribution, b) Pull distribution, c) Relative residual across pseudorapidity, and d) Relative residual across transverse momentum of the generator particle}
    \label{fig:calo_resolution_pion}
\end{figure}

The performance of the calorimeter calibration can be captured by inspecting the total energy of single particle samples. We use single electrons, photons and pions to study the resolution of the digitisation. Particles are generated with log-uniform energy between 300 MeV and 1 TeV. These sample sets are available as part of \colliderml{}. The total digitised energy deposited in the calorimeters is compared with the initial energy of the generator particle, producing either a residual $|E_{gen} - E_{calo}|$ or relative residual $(|E_{gen} - E_{calo}|) / E_{gen}$ or pull $(E_{gen} - E_{calo}) / \sqrt{E_{calo}}$ distribution. These are presented in \cref{fig:calo_resolution_electron}, \cref{fig:calo_resolution_photon} and \cref{fig:calo_resolution_pion}. Electron and photon energies are very well resolved by the digitisation, with less than 10\% residual at high $p_\text{T}$ and pseudorapidity between -2 and 2. This is consistent with the ceiling resolution of the ECal, as defined by its 3\% sampling fraction.

Pion energies are, as often seen in realistic detector descriptions, harder to resolve. Even at transverse momentum above 10 GeV, resolution plateaus at 30\%. In subsequent iterations of the dataset, this may be improved by dedicated tuning or, indeed, by machine learning-based calibration algorithms. 

It should be noted that not all initial state particles (that is, particles without parents in the dataset release) will enter the detector coverage. The majority of these are analogous to beam remnant particles with very high pseudorapidity. Of the 14TeV of kinetic energy involved in each $pp$ interaction, of order O(300)GeV will be deposited in the calorimeters, after applying the calibration described in \cref{tab:calo_calibration}. This is comparable with observations in LHC detectors \cite{Bayatian:2006jz,Cerci:2241148}.

\subsection{Object Data Types}

The following tables detail the exact data types and field names used in the dataset.

\begin{longtable}{p{4cm}p{3cm}p{7cm}}
\toprule
\textbf{Field} & \textbf{Type} & \textbf{Description} \\
\midrule
\endfirsthead
\toprule
\textbf{Field} & \textbf{Type} & \textbf{Description} \\
\midrule
\endhead
\texttt{event\_id} & \texttt{int32} & Unique event identifier \\
\texttt{particle\_id} & \texttt{list<uint64>} & Unique particle ID within event \\
\texttt{pdg\_id} & \texttt{list<int64>} & PDG particle code \\
\texttt{mass} & \texttt{list<float32>} & Particle rest mass (GeV/c²) \\
\texttt{energy} & \texttt{list<float32>} & Particle total energy (GeV) \\
\texttt{charge} & \texttt{list<float32>} & Electric charge (in units of $e$) \\
\texttt{px}, \texttt{py}, \texttt{pz} & \texttt{list<float32>} & Momentum components (GeV/$c$) \\
\texttt{vx}, \texttt{vy}, \texttt{vz} & \texttt{list<float32>} & Vertex position (mm) \\
\texttt{time} & \texttt{list<float32>} & Production time (ns) \\
\texttt{vertex\_primary} & \texttt{list<uint16>} & Primary vertex ID (1=hard scatter, [2,...,N]=pile-up) \\
\texttt{perigee\_d0} & \texttt{list<float32>} & Radial impact parameter of the particle's fitted perigee \\
\texttt{perigee\_z0} & \texttt{list<float32>} & Longitudinal impact parameter of the particle's fitted perigee \\
\texttt{parent\_id} & \texttt{list<int64>} & ID of parent particle \\
\texttt{primary} & \texttt{list<boolean>} & Whether the particle is primary or not \\
\bottomrule
\caption{Particle data structure. Note that \texttt{parent\_id} is given as $-1$ in case the particle has no parent.}\label{tab:particle_structure}
\end{longtable}

\begin{longtable}{p{4cm}p{3cm}p{7cm}}
\toprule
\textbf{Field} & \textbf{Type} & \textbf{Description} \\
\midrule
\endfirsthead
\toprule
\textbf{Field} & \textbf{Type} & \textbf{Description} \\
\midrule
\endhead
\texttt{event\_id} & \texttt{uint32} & Unique event identifier \\
\texttt{x}, \texttt{y}, \texttt{z} & \texttt{list<float32>} & Measured hit position (mm) \\
\texttt{true\_x}, \texttt{true\_y}, \texttt{true\_z} & \texttt{list<float32>} & True hit position before digitisation (mm) \\
\texttt{time} & \texttt{list<float32>} & Hit time (ns) \\
\texttt{particle\_id} & \texttt{list<uint64>} & Truth particle that created this hit \\
\texttt{volume\_id} & \texttt{list<uint8>} & Detector volume identifier \\
\texttt{layer\_id} & \texttt{list<uint16>} & Detector layer number \\
\texttt{surface\_id} & \texttt{list<uint32>} & Sensor surface identifier \\
\texttt{detector} & \texttt{list<uint8>} & Detector subsystem code \\
\bottomrule
\caption{Tracker hits data structure}\label{tab:tracker_hits_structure}
\end{longtable}

\begin{longtable}{p{4cm}p{3.5cm}p{6.5cm}}
\toprule
\textbf{Field} & \textbf{Type} & \textbf{Description} \\
\midrule
\endfirsthead
\toprule
\textbf{Field} & \textbf{Type} & \textbf{Description} \\
\midrule
\endhead
\texttt{event\_id} & \texttt{uint32} & Unique event identifier \\
\texttt{detector} & \texttt{list<uint8>} & Calorimeter subsystem name \\
\texttt{total\_energy} & \texttt{list<float32>} & Total energy deposited in cell (GeV) \\
\texttt{x}, \texttt{y}, \texttt{z} & \texttt{list<float32>} & Cell center position (mm) \\
\texttt{contrib\_particle\_ids} & \texttt{list<list<uint64>>} & IDs of particles contributing to this cell \\
\texttt{contrib\_energies} & \texttt{list<list<float32>>} & Energy contribution from each particle (GeV) \\
\texttt{contrib\_times} & \texttt{list<list<float32>>} & Time of each contribution (ns) \\
\bottomrule
\caption{Calorimeter hits data. Note the nested lists for contributions.}\label{tab:calo_hits_structure}
\end{longtable}

\begin{longtable}{p{4cm}p{3cm}p{7cm}}
\toprule
\textbf{Field} & \textbf{Type} & \textbf{Description} \\
\midrule
\endfirsthead
\toprule
\textbf{Field} & \textbf{Type} & \textbf{Description} \\
\midrule
\endhead
\texttt{event\_id} & \texttt{uint32} & Unique event identifier \\
\texttt{track\_id} & \texttt{list<uint16>} & Unique track identifier within event \\
\texttt{majority\_particle\_id} & \texttt{list<uint64>} & Truth particle with most hits on this track \\
\texttt{d0} & \texttt{list<float32>} & Transverse impact parameter (mm) \\
\texttt{z0} & \texttt{list<float32>} & Longitudinal impact parameter (mm) \\
\texttt{phi} & \texttt{list<float32>} & Azimuthal angle (radians) \\
\texttt{theta} & \texttt{list<float32>} & Polar angle (radians) \\
\texttt{qop} & \texttt{list<float32>} & Charge divided by momentum ($e$/GeV) \\
\texttt{hit\_ids} & \texttt{list<list<uint32>>} & List of tracker hit IDs assigned to this track \\
\bottomrule
\caption{Track data structure. Track parameters are those fitted to a helix by ACTS}\label{tab:tracks_structure}
\end{longtable}

\section{Public Collider Datasets}
\label{app:datasets}

The following datasets are included in \cref{fig:collider_datasets}:

\begin{table}[h!]
\centering
\small
\begin{threeparttable}
\caption{Public collider-physics datasets spanning different granularities of objects: Low (detector hits), Medium (tracks/calorimeter clusters), High (particle flow constituents), Analysis (fully reconstructed particles/tabular). Sizes are approximate as reported by sources. }
\label{tab:public-datasets}
\setlength{\tabcolsep}{3pt}
\renewcommand{\arraystretch}{1.2} % more vertical spacing
\begin{tabular}{@{}l l r l r@{}}
\toprule
\textbf{Dataset} & \textbf{Reference} & \multicolumn{1}{c}{\textbf{Samples (approx.)}} & \textbf{Level} & \multicolumn{1}{c}{\textbf{Size (approx.)}} \\
\midrule
LHC Olympics 2020 (LHCO2020) & \cite{lhco2020} & 4{,}000{,}000 & High & 11 GB \\ [3pt]
Dark Machines & \cite{darkmachines} &  O(1) billion & Analysis & 34 GB \\ [3pt]
COLLIDE-2V & \cite{moreno2025collide2v} & 750,000,000 & High & 50~TB \\
JetClass & \cite{jetclass,qu2024particletransformerjettagging} & 125{,}000{,}000 jets & High & 190 GB \\ [3pt]
ATLAS Top Tagging Open Data & \cite{atlastop} & 44{,}500{,}000 jets & High & 129 GB \\ [3pt]
JetNet30 & \cite{jetnet30} & \multirow{2}{*}{880{,}000 jets} & \multirow{2}{*}{High} & 440 MB \\ 
 JetNet150 & \cite{jetnet150} &  &  & 2.1 GB \\ [3pt]
TrackML & \cite{trackml_acc, trackml_through} & 10{,}000 & Low & 200 GB \\ [3pt]
\multirow{2}{*}{CaloChallenge 2022} & \cite{calochallenge_d2} (D2) & \multirow{2}{*}{200{,}000 showers} & \multirow{2}{*}{Low} & 2.7 GB \\
 & \cite{calochallenge_d3} (D3) &  &  & 7.6 GB \\ [3pt]
 TPCpp--10M & \cite{tpcpp10m} & 10{,}000{,}000 & Low & 120 GB \\ [3pt]
FAIR Higgs decay dataset & \cite{fairhiggs} & 6,000,000 jets & Medium & 228 GB \\[3pt]
CMS Open Data (2016) & \cite{cms_open} & O(10) billion & Medium & O(100) TB \\[3pt]
ATLAS Open Data (2015--2016) & \cite{atlas_open} & O(10) billion & Medium & O(100) TB \\
SUSY (UCI) & \cite{susy_uci} & 5{,}000{,}000 & Analysis & 880 MB \\ [3pt] 
\midrule
\multirow{2}{*}{ColliderML} & \multirow{2}{*}{This work} & \multirow{2}{*}{1{,}000{,}000} & \multirow{2}{*}{Low} & 400~TB (EDM4hep) \\
 & & & & 40~TB (Parquet) \\
\bottomrule
\end{tabular}
\end{threeparttable}
\end{table}

%%%%%%%%%%%%%%%%%%%%%%%%%%%%%%%%%%%%%%%%%%%%%%%%%%%%%%%%%%%%

\newpage

\end{document}